\documentclass[a4paper,fleqn,usenatbib,twocolumn]{mnras}

\usepackage{graphicx}	% Including figure files
\usepackage{amsmath}	% Advanced maths commands
\usepackage{amssymb}	% Extra maths symbols
\usepackage{savesym}
\usepackage{natbib}

\usepackage{mathrsfs,dsfont}
\usepackage{multirow}
\usepackage{cleveref}
\usepackage[skip=0pt]{caption}
\usepackage{float}
\usepackage{booktabs}
\usepackage{epstopdf}
\usepackage{romannum}
\usepackage{color}
\usepackage{hyperref}
\usepackage{verbatim}

\usepackage[none]{hyphenat}
\usepackage{soul}

\graphicspath{{figures/}}

\def\be{\begin{equation}}
\def\ee{\end{equation}}
\def\kms{{\rm \,km\,s^{-1}}}

\def\Gyr{{\rm \,Gyr}}

\def\kpc{{\rm \,kpc}}

\def\keV{{\rm \,keV}}

\def\msun{{\,M_\odot}}

\newcommand{\dd}{{\rm d}}
\newcommand{\Fr}{{\rm Fr}}
\newcommand{\chandra}{{\em Chandra}}

\defcitealias{Zhang2018}{ZCS18}

\usepackage{xcolor}
\title[Bubble-driven Gas Uplift in Galaxy Clusters]{Bubble-driven Gas Uplift in Galaxy Clusters and its Velocity Features}

\author[Congyao Zhang et al.]{
Congyao Zhang,$^{1}$\thanks{E-mail: cyzhang@astro.uchicago.edu}
Irina Zhuravleva,$^1$
Marie-Lou Gendron-Marsolais,$^{2,3}$
Eugene Churazov,$^{4,5}$
\newauthor
Alexander A. Schekochihin,$^{6,7}$
and
William R. Forman$^8$
\vspace{5pt}
\\
$^1$~Department of Astronomy and Astrophysics, The University of Chicago, Chicago, IL 60637, USA \\
$^2$~European Southern Observatory, Alonso de C\'{o}rdova 3107, Vitacura, Casilla 19001, Santiago de Chile \\
$^3$~Instituto de Astrof\'{i}sica de Andaluc\'{i}a (IAA-CSIC), Glorieta de la Astronom\'{i}a, 18008 Granada, Spain \\
$^4$~Max Planck Institute for Astrophysics, Karl-Schwarzschild-Str. 1, D-85741 Garching, Germany  \\
$^5$~Space Research Institute (IKI), Profsoyuznaya 84/32, Moscow 117997, Russia \\
$^6$~Rudolf Peierls Centre for Theoretical Physics, University of Oxford, Clarendon Laboratory, Parks Road, Oxford OX1 3PU, UK \\
$^7$~Merton College, Oxford OX1 4JD, UK \\
$^8$~Smithsonian Astrophysical Observatory, Harvard-Smithsonian Center for Astrophysics, 60 Garden St., Cambridge, MA 02138 \\
\vspace{-15pt}
}

\date{Accepted XXX. Received YYY; in original form ZZZ}

\pubyear{2022}

\begin{document}
\label{firstpage}
\pagerange{\pageref{firstpage}--\pageref{lastpage}}
\maketitle

\begin{abstract}

Buoyant bubbles of relativistic plasma are essential for active galactic nucleus feedback in galaxy clusters, stirring and heating the intracluster medium (ICM). Observations suggest that these rising bubbles maintain their integrity and sharp edges much longer than predicted by hydrodynamic simulations. In this study, we assume that bubbles can be modeled as rigid bodies and demonstrate that intact bubbles and their long-term interactions with the ambient ICM play an important role in shaping gas kinematics, forming thin gaseous structures (e.g., H$\alpha$ filaments), and generating internal waves in cluster cores. We find that well-developed eddies are formed in the wake of a buoyantly rising bubble, and it is these eddies, rather than the Darwin drift, that are responsible for most of the gas mass uplift. The eddies gradually elongate along the bubble's direction of motion due to the strong density stratification of the atmosphere and eventually detach from the bubble, quickly evolving into a high-speed jet-like stream propagating towards the cluster center in our model. This picture naturally explains the presence of long straight and horseshoe-shaped H$\alpha$ filaments in the Perseus cluster, inward and outward motions of the gas, and the X-ray-weighted gas velocity distributions near the northwestern bubble observed by \textit{Hitomi}. Our model reproduces the observed H$\alpha$ velocity structure function of filaments, providing a simple interpretation for its steep scaling and normalization: laminar gas flows and large eddies within filaments driven by the intact bubbles, rather than spatially homogeneous small-scale turbulence, are sufficient to produce a structure function consistent with observations.

\end{abstract}


%%%%%%%%%%%%%%%%%%%%%%%%%%%%%%%%%%%%%%%%%%%%%%%%%%
%%%%%%%%%%%%%%%%% BODY OF PAPER %%%%%%%%%%%%%%%%%%

\section{Introduction} \label{sec:introduction}

\begin{figure}
\centering
\includegraphics[width=1\linewidth]{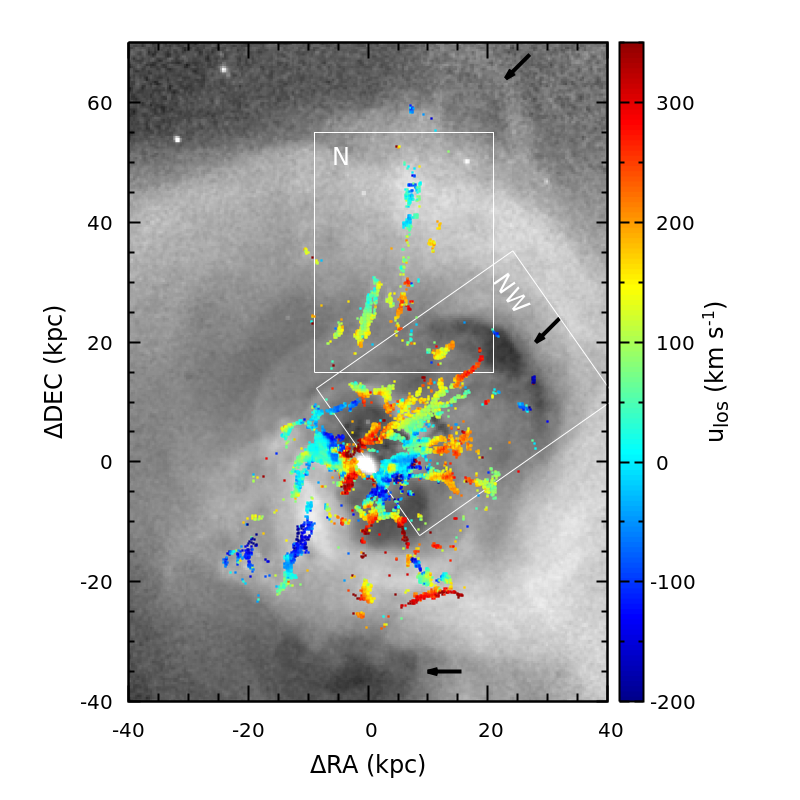}
\caption{Line-of-sight (LOS) velocity map of the H$\alpha$ filaments (only pixels with fitted H$\alpha$ flux higher than $4\times10^{-17}\,{\rm erg\,s^{-1}\,cm^{-2}\,pixel^{-1}}$ are shown; see \citealt{Gendron-Marsolais2018}) overlaid on the residual X-ray image (grey) of the Perseus cluster. The black arrows indicate three outer X-ray bubbles (or bubble candidates) tens of kpc from the cluster center. The filaments enclosed within the white rectangles are compared with our models in Fig.~\ref{fig:halpha_comp}. This figure shows the distribution of the multiphase gas in Perseus. The thin H$\alpha$ filaments are mostly elongated in the radial directions, associated with the X-ray bubbles (see Section~\ref{fig:perseus}).}
\label{fig:perseus}
\end{figure}

Co-existence of the multiple phases of gas, from the hot, weakly magnetized plasma (a.k.a., intracluster medium, ICM) to the cold ionized and molecular gas (see Fig.~\ref{fig:perseus} for the Perseus cluster as an example), in the cores of galaxy clusters, indicates an active competition between efficient radiative cooling and energetic heating processes (see, e.g., \citealt{McNamara2007,Fabian2012,Werner2019} for reviews).

Radio-mode active galactic nucleus (AGN) feedback has been recognized as a promising heating mechanism in cool cores of galaxy clusters based mainly on two facts. First, in observations, the central supermassive black holes are found to inject sufficient energy into inflated X-ray cavities (or bubbles) to balance the cooling loss from the inner ICM \citep[e.g.,][]{Churazov2000,McNamara2000,Birzan2004,Hlavacek-Larrondo2012}. Secondly, the energy-conservation law guarantees that the bubbles would eventually lose most of their energy to ambient gas atmospheres, independent of any specific energy-transfer mechanism \citep{Churazov2001,Churazov2002}. Multi-wavelength observations have shown clear evidence for interactions among the radio jets, bubbles, and their surrounding hot atmospheres \citep[e.g.,][]{Fabian2006,Forman2007,Tremblay2012,Sanders2016}.

Cold filamentary structures embedded in the ICM are commonly detected in active cores of nearby clusters through their emission lines (e.g., H$\alpha$ and CO; see \citealt{McDonald2010,McDonald2012,Olivares2019}). Their formation mechanism, however, is still poorly understood. In general, two major scenarios were invoked in the literature. The filaments might be formed (1) from the cold-gas precipitation due to the local thermal instabilities \citep[e.g.,][]{Gaspari2012,Sharma2012,Li2014}, which may also happen in radiatively cooling outflows as suggested by \citet{Qiu2020,Qiu2021}; (2) when rising bubbles entrain gas from the cold-gas reservoir near the cluster center \citep[e.g.,][]{Churazov2001,Fabian2003,Revaz2008}. \citet{McNamara2016} have also proposed a picture that combines both (1) and (2). The first possibility has been extensively explored with numerical simulations, in which AGN feedback maintains global thermal equilibrium in the atmosphere. The second one, however, has barely been examined, mainly because numerical modeling of bubbles faces serious problems.

In mesh-based Eulerian hydrodynamic simulations, the rising bubbles are susceptible to fluid instabilities (e.g., Rayleigh–Taylor and Kelvin-Helmholtz instabilities) that destroy the bubbles rapidly, on the bubble's sound-crossing timescale \citep[see, e.g.,][]{Reynolds2015}. This is in tension with the observations showing chains of rising bubbles that maintain their integrity and have sharp boundaries (see the Perseus cluster in Fig.~\ref{fig:perseus} and also many other examples in clusters/groups, e.g., M87/Virgo, Hydra~A, NGC~5813, and Nest200047 in \citealt{Forman2007,Wise2007,Randall2015,Brienza2021}). High viscosity of the ICM could stabilize the bubble surface \citep{Reynolds2005}, however, it was disfavored by recent \chandra{} observations \citep[e.g.,][]{Roediger2013,Ichinohe2017,Wang2018,Zhuravleva2019}. Magnetic fields may also resolve the issue \citep[e.g.,][]{Kaiser2005,Ruszkowski2007,Diehl2008,Candelaresi2020}. However, their configurations and relevant microphysics, which can strongly affect gas dynamics, are still unclear. \citet{Scannapieco2008} applied a subgrid-turbulence model and argued that instantaneous gas instabilities smear the bubble boundary with the surrounding medium, and mix the ambient medium with the bubble plasma which serves to stabilize the bubbles (see also \citealt{Bruggen2009}).

The destruction of bubbles in simulations eliminates the important interactions between observed, long-lived bubbles and their environments, and hence biases our understanding of how bubbles stir and heat the ICM. In particular, it is vital to preserve the bubble integrity when modeling entrainment of gas by the bubbles in their wakes. Though bubbles tend to be much more stable in  smoothed-particle hydrodynamic (SPH) simulations \citep{Revaz2008} due to the fact that SPH codes induce strong numerical surface tension at the bubble surfaces \citep{Agertz2007}, the fluid behavior in the bubble wakes could not be well captured by the SPH method \citep[e.g.,][]{Wadsley2008,Bauer2012}.

\citet[hereafter \citetalias{Zhang2018}]{Zhang2018} proposed a rigid-bubble model in the mesh-based simulations to overcome the issue of bubble integrity: in their model, bubbles were assumed to experience no deformation during their buoyant rise. Despite such a strong assumption, the model captured several key features that had been missed in previous AGN feedback studies, including the excitation of internal gravity waves and well-developed wakes of the buoyant bubbles. The model implied that long-lived intact bubbles could dramatically change our view of how AGN feedback works in galaxy clusters, which motivates this project.

In this study, we use a similar numerical model as in \citetalias{Zhang2018} to explore the gas uplift by rising bubbles and the formation of filamentary structures during this process. A classical picture that describes the displacement of fluid parcels by a moving solid body, Darwin drift \citep{Darwin1953,Lighthill1956}, is only applicable to potential flow without stratification. The situation in galaxy cluster cores is more complicated, in part due to the presence of strong stratification. We find that most of the gas mass uplifted by bubbles in our simulations is uplifted through ``eddy transport'', rather than by Darwin drift (see Section~\ref{sec:result:pic} and also \citealt{Pope2010}).

The kinematics of the cold filaments could provide an independent probe of the gas velocity field of the ICM. \citet{Fabian2003} made a direct comparison between the morphology of the horseshoe-shaped filaments in the Perseus cluster and the streamlines formed in the wake of an air bubble rising in water (see their fig.~3), and used their similarity to argue that bubbles in the ICM drag up H$\alpha$ gas behind them. Our model numerically confirms this picture. In addition to that, we find that the evolution of the bubble wake is remarkably affected by the gravitational field, mainly characterized by the bubble's Froude number. Besides the characteristic horseshoe shape, the bubble-driven gas uplift is also sufficient to explain the formation of $\sim100\kpc$-long filaments (see Fig.~\ref{fig:perseus}). Filament velocities are naturally determined by the bubble's terminal velocity in our model.

Recent optical/sub-mm observations with unprecedented resolutions allow a more quantitative investigation of the filaments' velocity distribution \citep[e.g.,][]{Werner2013,Gendron-Marsolais2018,Russell2019}. \citet{Li2020} estimated the velocity structure function of H$\alpha$ filaments in nearby clusters and attributed it to AGN-driven turbulence. However, they found a steep scaling that did not follow the Kolmogorov five-thirds law. Though the presence of magnetic fields or supersonic turbulence may steepen the scaling \citep{Wang2021,Mohapatra2022,Hu2022}, our model provides an alternative, simpler explanation for the observational results. Namely, to account for the observed structure function, it is sufficient to assume that the filaments' velocities are dominated by laminar gas flows and large eddies formed during the bubble-driven gas uplift, rather than by uniformly-distributed small-scale turbulence. Therefore, the filament structure function cannot be used to constrain the properties of small-scale turbulence in the ICM (see Section~\ref{sec:obs:halpha}). The overlap of filaments on small scales and their sparse distribution on large scales also have strong effects on the structure function.

This paper is organized as follows. Section~\ref{sec:sim} describes the model and simulation method adopted in this work. In Section~\ref{sec:result}, we present the main results of our simulations, including the bubble-driven gas uplift and its dependence on the bubble parameters (i.e., Froude number, shape, and size). In Section~\ref{sec:obs}, we predict the observational signatures of the characteristic gas velocity pattern in the bubble wakes, including X-ray line broadening and LOS velocity of H$\alpha$ filaments. In Section~\ref{sec:conclusions}, we summarize our conclusions.

\section{Modeling Bubbles and their Uplifted Gas} \label{sec:sim}

We simulate bubbles rising in the ICM in the framework of the rigid-bubble model developed in \citetalias{Zhang2018}. By design, the bubbles maintain their integrity as they rise, providing a unique opportunity to investigate how long-term interactions between the bubble and the ambient atmosphere redistribute the ICM in a cluster core. Passive Lagrangian particles are included in the simulations to trace the gas motion driven by the bubbles.

\subsection{Model and simulation methods} \label{sec:sim:method}

\begin{figure}
\centering
\includegraphics[width=1\linewidth]{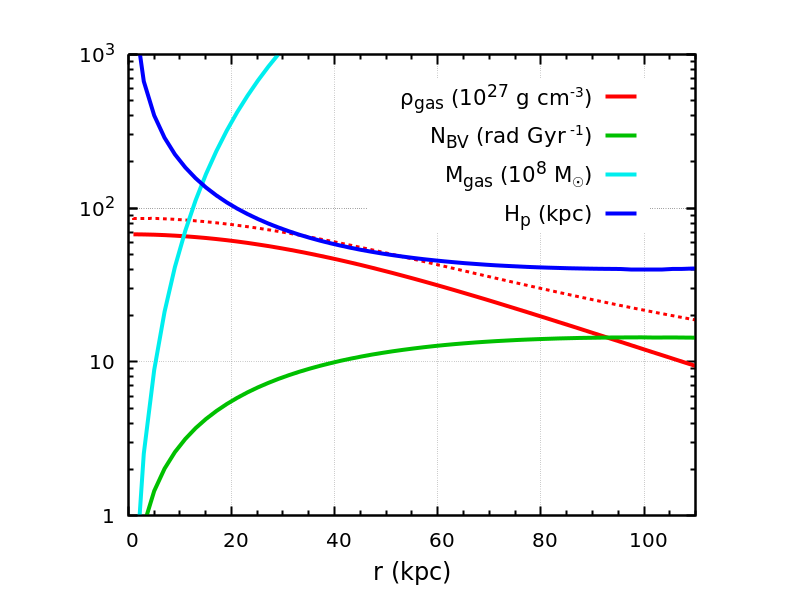}
\caption{Initial radial profiles of gas density (solid red line), Brunt–V\"{a}is\"{a}l\"{a} frequency (green line), enclosed gas mass (cyan line), and pressure scale height (blue line) used in our simulations. As a comparison, an analytical approximation for the radial density profile of the Perseus cluster is shown as the dotted red line (see Section~\ref{sec:sim:method}). }
\label{fig:init_profs}
\end{figure}

In all our simulations, we assume a static gravitational potential in spherical symmetry to model a cluster environment,
\be
\Phi(r) = 2V_{\rm c}^2\ln{\left[\Big(\frac{r}{R_{\rm core}}\Big)^2+1\right]},
\label{eq:potential}
\ee
where $V_{\rm c}=10^3\kms$ and $R_{\rm core}=10^2\kpc$ are the scaling parameters for the potential and the core radius, respectively. The hot gas within this atmosphere is initially isothermal and in hydrostatic equilibrium. The adiabatic index of the gas is $\gamma=5/3$. The corresponding gas density profile is, thus,
\be
\rho_{\rm gas}(r) = \rho_{\rm c}\exp\Big[-\frac{\Phi(r)}{c_{\rm t}^2}\Big],
\label{eq:density_profile}
\ee
where $\rho_{\rm c}=6.77\times10^{-26}{\rm\,g\,cm^{-3}}$ and $c_{\rm t}=\sqrt{k_{\rm B}T_{\rm gas}/\mu m_{\rm p}}$ are the central density and isothermal sound speed of the atmosphere; $T_{\rm gas},\ k_{\rm B},\ \mu\,(=0.6)$, and $m_{\rm p}$ are the initial gas temperature, Boltzmann constant, mean
molecular weight per ion, and proton mass, respectively. The gas temperature $k_{\rm B}T_{\rm gas}$ is fixed at $5\keV$ in our simulations. Note that we do not consider gas self-gravity in our model. The parameters used in Equations~(\ref{eq:potential}) and (\ref{eq:density_profile}) are selected so that our density profile is similar to that of the Perseus cluster within $\simeq10^2\kpc$. Fig.~\ref{fig:init_profs} shows the initial radial profiles of the gas density, enclosed gas mass $M_{\rm gas}$, pressure scale height $H_{\rm p}$, and Brunt–V\"{a}is\"{a}l\"{a} frequency $N_{\rm BV}\,(=\sqrt{(1-1/\gamma)}c_{\rm t}/H_{\rm p})$ of the atmosphere. For comparison, an analytical approximation for Perseus's gas density profile is shown as the dotted red line in Fig.~\ref{fig:init_profs}. We assume $\rho_{\rm gas}(r)=1.2n_{\rm e}(r)m_{\rm p}$, where $n_{\rm e}(r)$ is the best-fit electron number density profile given in \citet[][see their equation~4]{Churazov2003}.

The simulations are performed in a two-dimensional (2D) axisymmetric coordinate system $(x,\ y)$, also known as the 2.5D simulations, whose symmetric axis is along the $y$-axis. For convenience, we also define the $z$-direction by the right-hand rule in our model. For all our simulations, the computational domain is set to be $x\in[0,\ 50\kpc]$ and $y\in[-150,\ 200\kpc]$, which is sufficiently large for the boundaries of the simulation box to not affect our results. The effective resolution of our simulations reaches $0.5\kpc$. We have checked the numerical convergence by testing different resolutions, concluding that the simulation results are not affected (see Appendix~\ref{sec:appendix} for more discussions).

\begin{table*}
\centering
\begin{minipage}{0.9\linewidth}
\centering
\caption{Parameters of simulations (see Section~\ref{sec:sim:method}).}
\label{tab:sim_params}
\begin{tabular}{cccccc}
  \hline
  IDs\footnote{The IDs of our simulation runs, which are written as E$a$L$b$U$c$, where $a$ and $b$ indicate the bubble aspect ratio and horizontal size, $c$ represents the bubble velocity (this parameter is absent when bubble buoyantly rises in the atmosphere). } &
  $\varepsilon_{\rm bub}$\footnote{The bubble aspect ratio $\varepsilon_{\rm bub}\,(\equiv L_{\rm bub}/h_{\rm bub})$.}  &
  $L_{\rm bub}\,(\rm kpc)$\footnote{The bubble width $L_{\rm bub}$, i.e., the length scale along the direction perpendicular to the bubble velocity.}  &
  $U_0\,(\rm km\,s^{-1})$\footnote{The constant bubble velocity if applicable.}  &
  Motion\footnote{The type of bubble motion.} &
  Sampling\footnote{The type of sampling strategy for the Lagrangian tracer particles.}

\\ \hline
  E2L12         & $2$    & 12   & --    & float     & type~I  \\
  E4L12         & $4$    & 12   & --    & float     & type~I  \\
  E4L12S        & $4$    & 12   & --    & float     & type~II \\
  E4L24         & $4$    & 24   & --    & float     & type~I  \\
  E8L12         & $8$    & 12   & --    & float     & type~I  \\
  E4L12U75      & $4$    & 12   & 75    & constant  & type~I  \\
  E4L12U150     & $4$    & 12   & 150   & constant  & type~I  \\
  E4L12U300     & $4$    & 12   & 300   & constant  & type~I  \\
  E4L6U75       & $4$    & 6    & 75    & constant  & type~I  \\
  E4L24U300     & $4$    & 24   & 300   & constant  & type~I  \\
\hline
\vspace{-7mm}
\end{tabular}
\end{minipage}
\end{table*}

In each simulation, a rigid bubble has the shape of a spherical cap \citep{Gull1973} and is modeled as a wall with a slip boundary condition (i.e., allowing the gas to move along the boundary). Such a shape is motivated by both X-ray bubbles observed in nearby clusters (e.g., the northwestern bubble in Perseus; see Fig.~\ref{fig:perseus}) and more general studies of gas bubbles moving in a liquid\footnote{In fact, surface tension at the bubble interface helps shape the bubble morphology in this situation.} \citep[e.g.,][]{Bhaga1981,Tripathi2015}. The bubble's maximum width and height are denoted as $L_{\rm bub}$ and $h_{\rm bub}$, respectively. In this study, we explore the parameter space of the bubbles, including their width, aspect ratio $\varepsilon_{\rm bub}\,(\equiv L_{\rm bub}/h_{\rm bub})$, and rise velocity (see Table~\ref{tab:sim_params} for a summary). The sizes of our bubbles are, however, always smaller than the pressure scale height of the atmosphere (see Fig.~\ref{fig:init_profs}), as is the case in the Perseus cluster (where $H_{\rm p}\simeq30-50\kpc$ at $r\simeq10-30\kpc$). In all our simulations, the bubbles are initially static and their bottom boundaries are located at $y=5\kpc$. Note that our model skips the bubble's early rapid inflation phase caused by AGN-driven jet/outflow but only focuses on the stage when the bubble has already detached from the cluster center and approached its terminal velocity. We modeled two types of bubble motions: (1) moving with a constant velocity $U_0$\footnote{In this case, the bubble is given a rapid constant acceleration of $10^4\kms\Gyr^{-1}$ until its bubble velocity reaches $U_0$. This way the bubble avoids a sudden velocity jump at the start of the simulation, which would cause numerical difficulties.} and (2) rising buoyantly in a stratified atmosphere (see Section~\ref{sec:sim:u_bub} for the evolution of bubble-rise velocities). The bubble acceleration is determined by $F_{\rm bub}/M_{\rm bub}$, where $F_{\rm bub}$ is the pressure and viscous force acting on the bubble surface along the $y$-axis (see equation~10 in \citetalias{Zhang2018}), $M_{\rm bub}$ is the bubble's inertial mass determined by the uniform bubble density ($8\times10^{-28}\,{\rm g\,cm^{-3}}$, smaller than $0.1\rho_{\rm gas}$). The gravitational force acting on the bubble is ignored in the simulation.

Our simulations are performed with the open source mesh-based code OpenFOAM.\footnote{Open Source Field Operation and Manipulation, version v2106, \href{https://www.openfoam.com}{www.openfoam.com}. } We modified the built-in solver rhoPimpleFoam to solve the compressible fluid dynamics in a static gravitational field with Lagrangian tracer particles. The one-equation eddy-viscosity model (kEqn) is employed to handle the subgrid turbulence. We set the dynamic viscosity of the atmosphere as $0.3\rm\,g\,cm^{-1}\,s^{-1}$, smaller than 2 per cent of the Spitzer value \citep{Braginskii1958,Spitzer1962}. However, we emphasize that small-scale turbulence could not be properly captured in our 2.5D simulations by design. More details of the simulation method can be found in \citetalias{Zhang2018} (see their appendix).

We applied two strategies to sample Lagrangian tracer particles in our simulations. One is to distribute the particles uniformly inside the computational domain within $r<50\kpc$ (type~I). In this way, the gas flow near the axis of symmetry could be well resolved for the purpose of visualizations (i.e., having a higher mass resolution; see Figs.~\labelcref{fig:part_a4vb_1,fig:part_a4vb_2} for examples). The other strategy is to sample the particles depending on the gas mass distribution of the atmosphere, assuming each particle traces the same amount of gas mass (type~II). It provides an unbiased particle distribution and is convenient for quantifying the gas flow with these particles (see Figs.~\labelcref{fig:fpart_disp,fig:traj}). In OpenFOAM, the Lagrangian particles are assumed to be spheres with radius $r_{\rm part}$ and uniform density $\rho_{\rm part}$. Their motions are determined by the drag force $C_{\rm D}\pi r_{\rm part}^2\Delta u^2/2$ and particle mass $4\pi r_{\rm part}^3\rho_{\rm part}/3$, where $C_{\rm D}$ is the drag coefficient ($\simeq0.5-1$ in our relevant Reynolds number regime; see \citealt{Schiller1935}) and $\Delta u$ is the particle velocity relative to its ambient gas. We set a sufficiently small particle radius ($r_{\rm part}=10^{-2}\kpc$) and density ($\rho_{\rm part}=10^{-28}\,{\rm g\,cm^{-3}}$) for all our particles to couple tightly with the atmosphere. We have tested a wide range of these parameters. Our results show little dependence on them.

\subsection{Rise velocity of buoyant bubbles} \label{sec:sim:u_bub}

\begin{figure}
\centering
\includegraphics[width=1\linewidth]{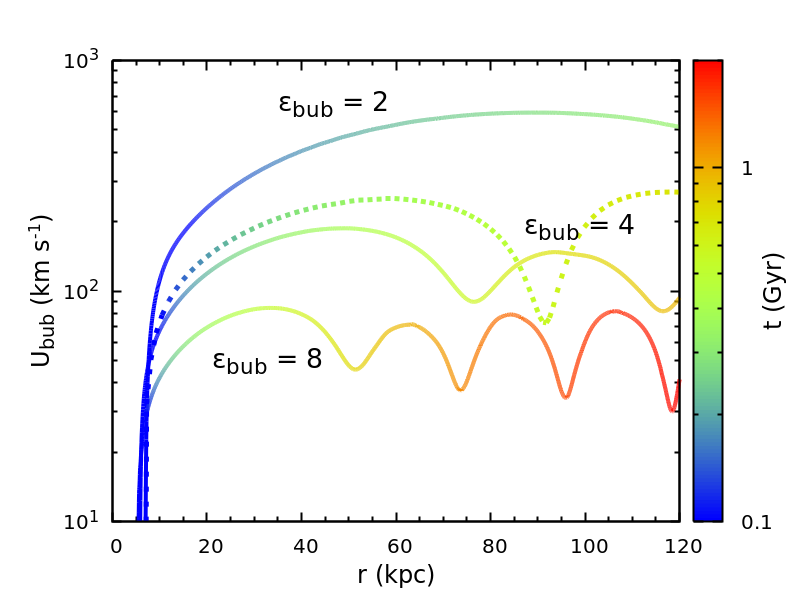}
\caption{Evolution of the bubble-rise velocities in the simulations E2L12, E4L12, E8L12 (solid lines), and E4L24 (dotted line). The horizontal axis represents the radial position of the bubble center. The line color encodes the simulation time. This figure shows that our bubbles quickly approach their terminal velocities ($t\lesssim0.3\Gyr$) when buoyantly rising in the gravitationally stratified atmosphere. The terminal velocity depends on both bubble size and shape (see Section~\ref{sec:sim:u_bub}). }
\label{fig:uterm}
\end{figure}

Fig.~\ref{fig:uterm} shows the evolution of the bubble velocity $U_{\rm bub}$ while bubbles rise buoyantly in the stratified atmosphere. If $L_{\rm bub}$ is fixed, the flatter bubbles (larger $\varepsilon_{\rm bub}$) have smaller terminal velocities, consistent with those presented in \citetalias{Zhang2018}. The oscillations of the velocity curves are caused by vortex shedding -- periodic detachments of downstream eddies from the bubble surface, which dramatically change the gas velocity and pressure distributions around the bubble. We find that the oscillation period is $T_{\rm shed}\simeq0.5\Gyr$, only mildly dependent on the bubble's aspect ratio and size. Due to strong stratification, the radial length scale in the bubble wake is set by $\rm Fr=1$, namely, $\ell_{\parallel}=U_{\rm bub}/N_{\rm BV}$. The evolution of any structures radially larger than $\ell_{\parallel}$ is affected significantly by the gravitational field (see more discussion in \citetalias{Zhang2018}). Thus, the dimensionless Strouhal number, commonly used to characterize oscillations of the fluid, can be estimated as
\be
{\rm St} = \frac{\ell_{\parallel}}{T_{\rm shed}U_{\rm bub}} \sim 0.2.
\ee
It is close to low-frequency-mode St driven by large-scale instabilities of the wake, e.g., vortex shedding, broadly reported in the literature \citep[e.g.,][]{Sakamoto1990,Nakamura1996}. In Section~\ref{sec:result}, we will mostly focus on the first period of the oscillations, when the gas uplift from the cluster center occurs.

\section{Gas Uplift by Bubbles} \label{sec:result}

In this section, we explore how gas in the cluster core is disturbed by rising bubbles. The entire process could be generally summarized as a two-stage scenario -- uplift and detachment (see Section~\ref{sec:result:pic}). The bubble's Froude number is the fundamental parameter controlling the evolution of the entire system (see Sections~\labelcref{sec:result:fr,sec:result:shape}).

\subsection{A general picture} \label{sec:result:pic}

\begin{figure*}
\centering
\includegraphics[width=0.9\linewidth]{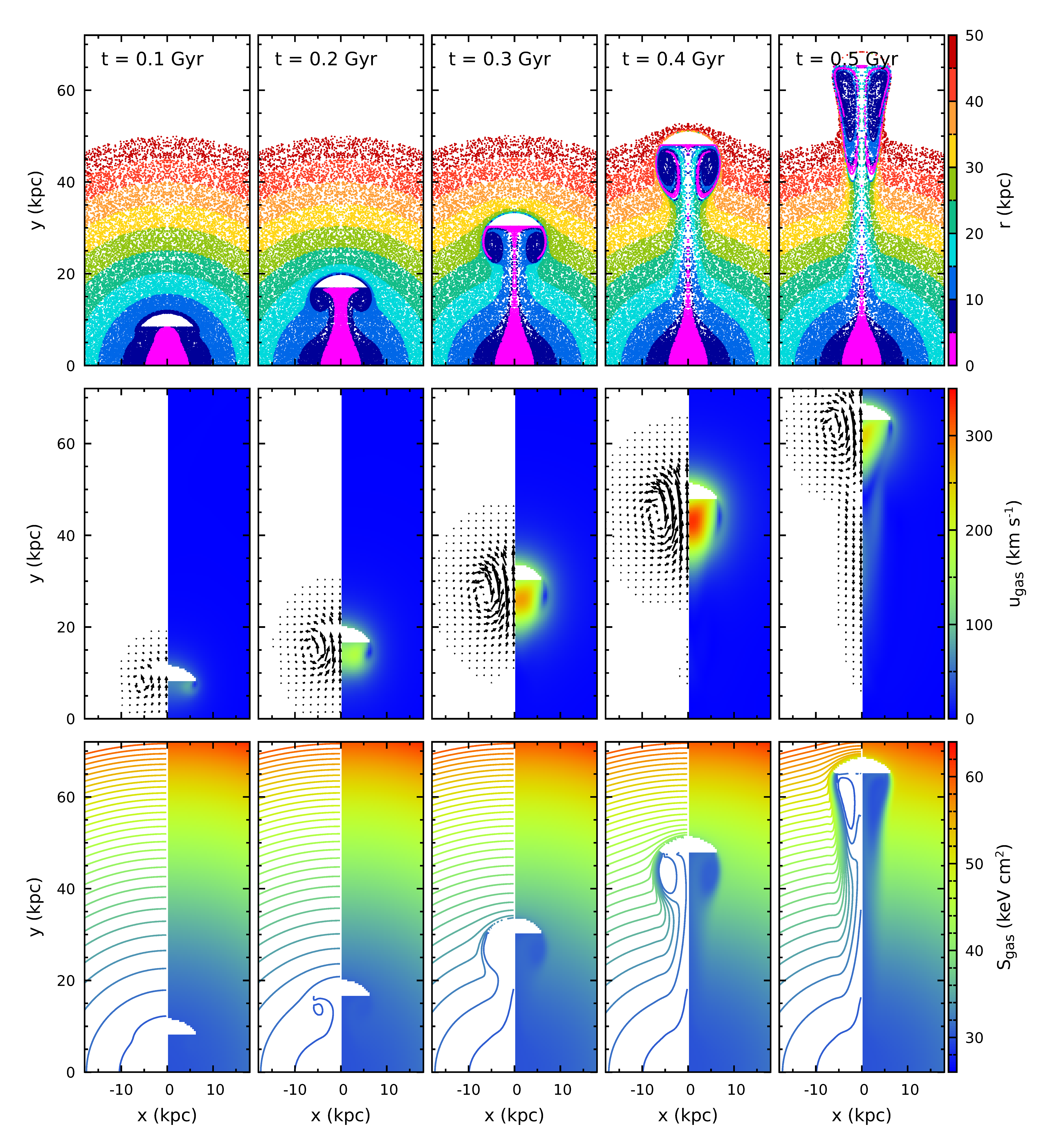}
\caption{\textit{Top panels:} Distributions of the Lagrangian tracer particles during the evolution of a buoyantly rising bubble (with spherical cap shape) in the E4L12 simulation. The color encodes their initial radial positions. The images are mirrored across the $y$-axis ($x=0$). \textit{Middle panels:} Corresponding gas velocity fields in the rest frame of the cluster. The left halves of the panels show the velocity vector fields; the right halves show the total gas velocity. \textit{Bottom panels:} Evolution of the gas entropy. The interval between each two successive contour levels is $1\rm\,keV\,cm^{2}$. This figure illustrates how a buoyant bubble uplifts gas from the cluster center. Eddies (see, e.g., dark blue particles trailing the bubble in the top panels) are quickly formed downstream of the rising bubble, where the gas velocity can be up to $\simeq2$ times larger than the bubble-rise velocity (see Section~\ref{sec:result:pic}).}
\label{fig:part_a4vb_1}
\end{figure*}

Fig.~\ref{fig:part_a4vb_1} illustrates how the ICM in a cluster core is uplifted by a buoyant bubble in our simulation E4L12. The top panels show the distributions of tracer particles colored based on their initial radial positions. The images are mirrored across the $y$-axis. The middle and bottom panels show the corresponding gas velocity and entropy fields. In the rest frame of the cluster, the particles just ahead of the bubble are pushed away from the bubble's path and most of them end up in the bubble wake. Those particles initially behind the bubble gain velocity due to the pressure gradient generated in the wake and rise upwards. Their distributions are prominently elongated along the radial direction (see, e.g., the purple particles).

Downstream vortices (or eddies) appear in the bubble wake shortly after the start of the simulation. They continuously suck in gas from the rear, near the centerline of the bubble, clearly seen in the velocity vector fields. Those eddies develop on the timescale $L_{\rm bub}/U_{\rm bub}$ and are filled with the ICM largely from the inner cluster region. For instance, one can see that, at $t\simeq0.4\Gyr$, the eddies are mostly made up by the gas initially residing at $r\lesssim10\kpc$. Once flowing into the eddies, the gas parcels are stretched (see the purple particles in Fig.~\ref{fig:part_a4vb_1}, and also in Fig.~\ref{fig:part_a4l}), forming thin gaseous structures (similar to thin H$\alpha$ filaments). The gas-entropy distributions show similar results to the tracer particles. The gas velocity in the eddies can be up to a factor of $\simeq2$ higher than the bubble-rise velocity and could leave imprints in the high-resolution X-ray spectra of the ICM (see Section~\ref{sec:obs:xray}).

\begin{figure}
\centering
\includegraphics[width=1\linewidth]{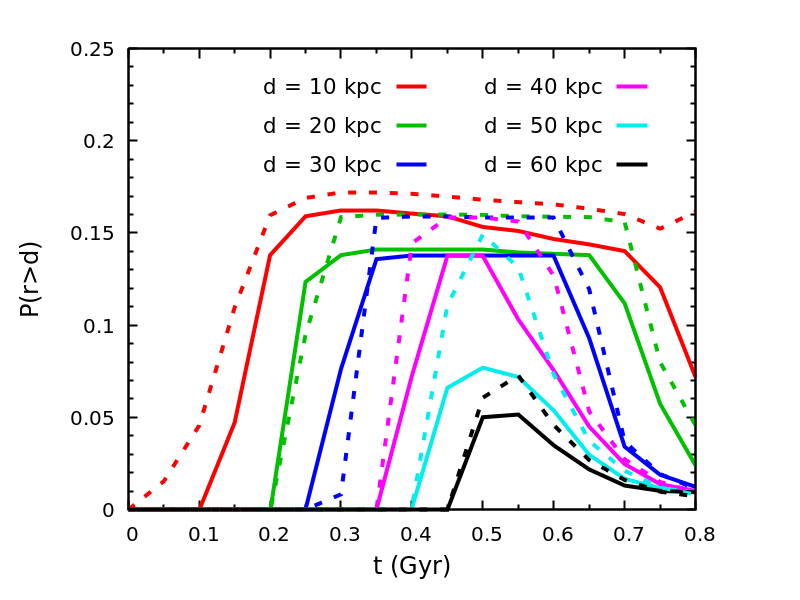}
\caption{Evolution of the number fractions of the tracer particles uplifted to radii larger than $d$ (color coded in the legend) in our simulation E4L12S. The solid and dashed lines show the fractions of particles that initially reside in the regions $r<5\kpc$ and $5<r<10\kpc$, respectively. The flat and equal tops of the green/blue/purple curves show the dominant role of the eddy transport in lifting the cluster inner gas up to a large radius (e.g., $\gtrsim10\kpc$; see Section~\ref{sec:result:pic}).}
\label{fig:fpart_disp}
\end{figure}

Fig.~\ref{fig:fpart_disp} shows the number fractions of the tracer particles initially inside the regions $r<5\kpc$ (solid lines) and $5<r<10\kpc$ (dashed lines) that are uplifted to at least the radius $d$ in the simulation E4L12S. These number fractions also reflect the gas mass fractions because the particles trace the equal-mass gas parcels in this run. The hat-shaped green, blue, and purple lines ($d=20,\,30,\,40\kpc$) show almost the same peak fraction, indicating the fraction of particles inside and moving together with the eddies. About $5\times10^8\msun$ gas mass is uplifted this way by a single bubble, a few times larger than the gas mass displaced by the bubble volume near the cluster center (given $\simeq5\times10^9\msun$ total gas mass is within $r=10\kpc$ in our cluster; see Fig.~\ref{fig:init_profs}). In contrast, the red line ($d=10\kpc$) shows the fraction of the uplifted particles caused by both the Darwin drift and eddy transport. Comparing the red with green/blue/purple curves, we find that $\sim90$ per cent of the gas uplifted beyond $\simeq 10\kpc$ is uplifted by the eddies. This result demonstrates that the large fraction of the entrained gas is trapped inside the downstream eddies and moves together with the bubble, in contrast to the picture of the Darwin drift \citep{Darwin1953,Duan2018}. This highlights the importance of the non-linear effect, which was sketched as ``wake transport'' in \citet{Pope2010}. \citet{Pope2010} used a parameter $q$ to quantify such a fraction in their analytical model (see their equation~12), which approaches unity in our simulations. The Darwin drift contributes negligibly to the mass fraction that is uplifted beyond $r\gtrsim20\kpc$. It is worth noting that the quantitative results reported here may depend on the size, shape, and initial radial position of the bubble, but the general picture of the evolution does not change with the parameters explored in this study (see Section~\ref{sec:result:shape} for more discussions).

\begin{figure*}
\centering
\includegraphics[width=0.9\linewidth]{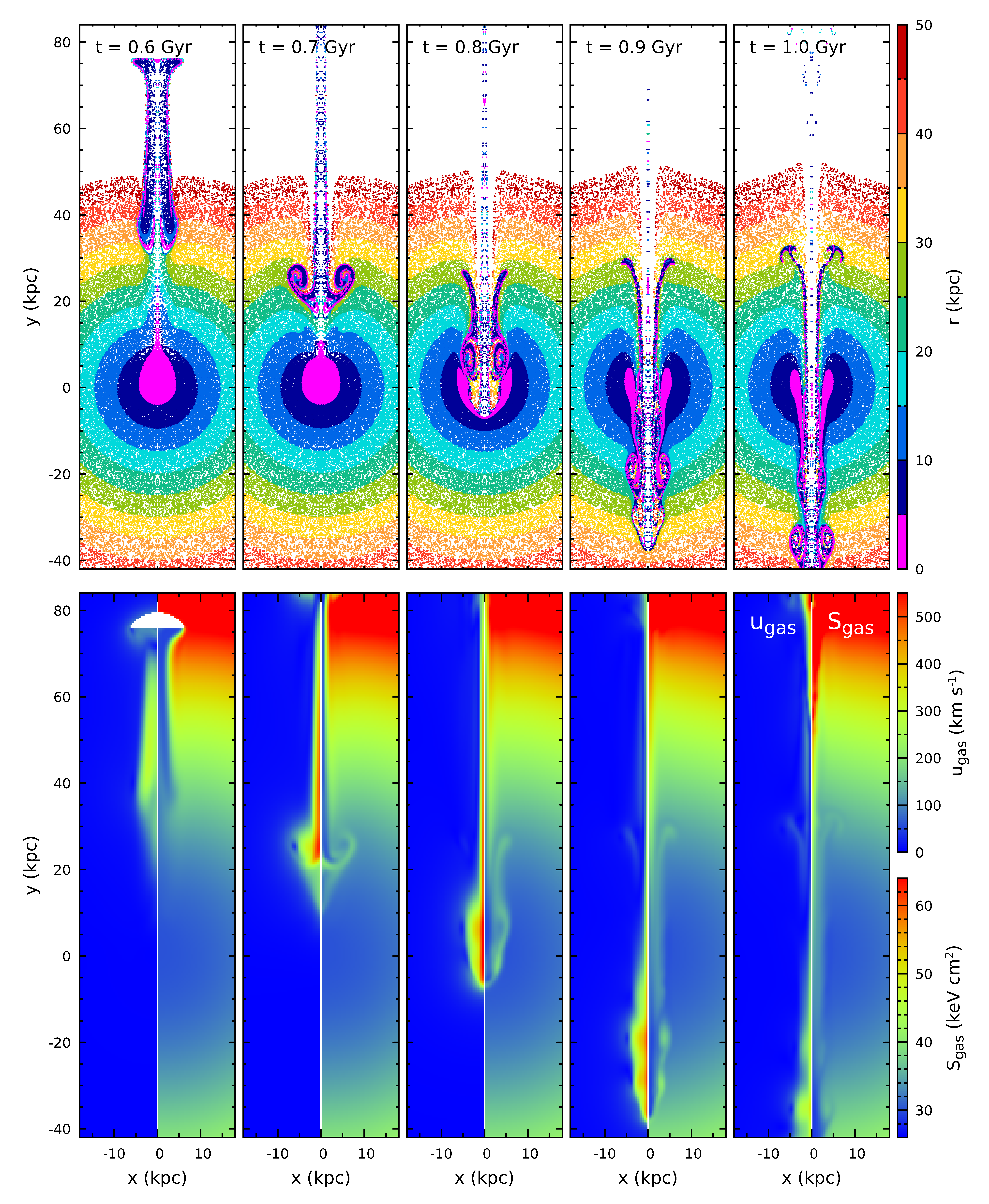}
\caption{Similar to Fig.~\ref{fig:part_a4vb_1}, but for later stages of the evolution, after the eddy detachment in the simulation E4L12. \textit{Top panels:} Distributions of the tracer particles. \textit{Bottom panels:} The left halves of the panels show the gas velocity distribution; the right halves show the corresponding gas entropy. This figure shows the formation of a reverse jet from the stretched eddies detaching from the bubble. This jet dramatically disturbs the gas in the cluster core (see Section~\ref{sec:result:pic}). }
\label{fig:part_a4vb_2}
\end{figure*}

Gravity dramatically alters the morphology of the bubble eddies when their radial size becomes comparable to the buoyancy length scale of the system $\ell_{\parallel}\,(=U_{\rm bub}/N_{\rm BV})$. In Fig.~\ref{fig:part_a4vb_1}, the eddies are gradually elongated along the bubble's direction of motion (radial) and shrunk in the azimuthal direction. They eventually detach from the bubble, e.g., starting at $t\simeq0.4\Gyr$ in the example shown in Fig.~\ref{fig:part_a4vb_1}. Their subsequent evolution after $t=0.6\Gyr$ is shown in Fig.~\ref{fig:part_a4vb_2}. The eddies are further stretched and quickly evolve into a high-speed reverse jet-like stream (referred to as jet hereafter) propagating towards the cluster center. Their central gas velocity could reach up to $\simeq3U_{\rm bub}$. A similar structure has been observed in \citetalias{Zhang2018} (see also, e.g., \citealt{Torres2000,Okino2021}). Strong Rayleigh–Taylor instabilities develop near the tip of the jet and trigger the formation of new large vortices. However, we note that the evolution of the high-speed jet flow may depend on both the Reynolds and Froude numbers of the system (\citealt{Magnaudet2020}; see their fig.~2). We defer a systematic study of these dependences to our future work.

\begin{figure}
\centering
\includegraphics[width=1\linewidth]{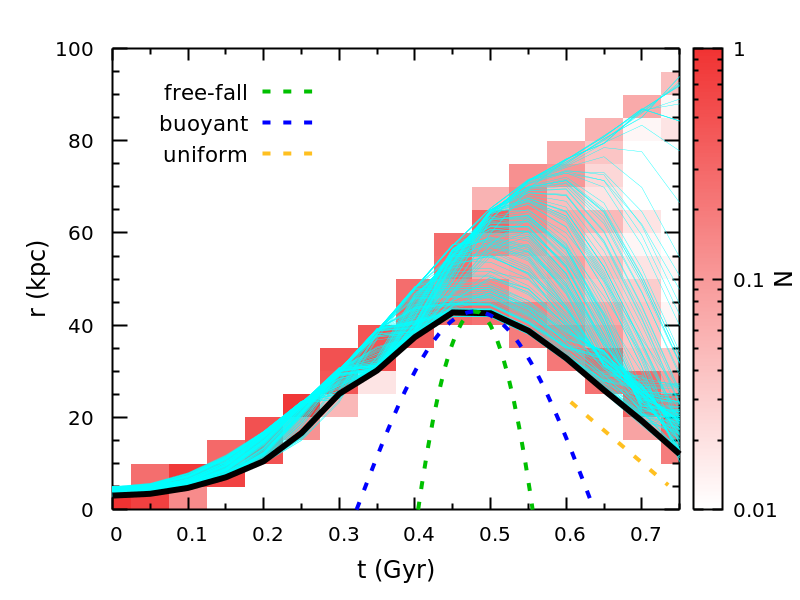}
\caption{Radial trajectories (cyan solid lines) of $163$ particles uplifted through eddy transport in the simulation E4L12S. These particles are initially within $r=5\kpc$ and relocate to $r>10\kpc$ at $t=0.75\Gyr$. Their normalized number distribution vs. $r$ is shown in the reddish color in the background. The thick black line marks the 5th percentile of these particles' radial positions, well tracing the radial position of the tip of the reverse jet (see Fig.~\ref{fig:part_a4vb_2}). For comparison, the trajectories representing three basic modes of gas motion are overlaid as the dashed lines, including free fall (green), buoyant oscillation (blue), and uniform motion ($v_{\rm d}=-140\kms$, yellow). This figure illustrates the formation of the reverse jet after the eddy detachment (near $t\simeq0.5\Gyr$). The particle distribution is stretched radially. Due to the ram pressure, the tip of the jet reaches its terminal velocity ($\simeq v_{\rm d}$), comparable to the bubble-rise velocity (see Section~\ref{sec:result:pic}).}
\label{fig:traj}
\end{figure}

Fig.~\ref{fig:traj} shows the radial trajectories of $163$ particles (cyan lines), all experiencing eddy transport, in the run E4L12. These particles are initially at $r\leq5\kpc$ (i.e., purple ones in Figs.~\labelcref{fig:part_a4vb_1,fig:part_a4vb_2}) and located at $r>10\kpc$ at $t=0.75\Gyr$. The solid black line marks the radial position smaller than $95$ per cent of the radial positions of these particles, well tracing the location of the tip of the reverse jet. It shows clearly the detachment of the eddies occurring around $t_{\rm d}=0.48\Gyr$. After that, the spatial distribution of the particles spreads radially. This timescale is comparable to the radiative cooling time of the uplifted dense and low-entropy gas from the cluster center. Thus, cold blobs of gas can be formed in the bubble wake. These blobs will not fall back immediately as they will follow the rising bubble. As they move up, their required time to fall to equilibrium locations will increase and become longer than the cooling time. This supports the scenario discussed in \citet{McNamara2016}, namely, the effect of enhancing thermal instabilities in the bubble-uplifted gas.

At $t\simeq0.7\Gyr$, a small fraction of the particles still follow the buoyant bubble. Most of others fall back towards the cluster center and pile up near the jet tip. To gain a better understanding of the particles' dynamical behavior, we overlay three basic modes of gas motion as dashed lines in Fig.~\ref{fig:traj}, including
\begin{enumerate}
  \item free fall (green): $r=r_{\rm d}-g(t-t_{\rm d})^2/2$,
  \item buoyant oscillation (blue): $r=r_{\rm d}\cos[N_{\rm BV}(t-t_{\rm d})]$, and
  \item uniform motion with constant velocity $v_{\rm d}=-140\kms$ (yellow),
\end{enumerate}
where $g=1.5\times10^{4}\,{\rm kpc\,Gyr^{-2}}$ and $N_{\rm BV}=10\,{\rm rad\,Gyr^{-1}}$ are typical gravitational acceleration and Brunt–V\"{a}is\"{a}l\"{a} frequency based on the initial density and pressure profiles of our cluster; $r_{\rm d}=43\kpc$ is the maximum radius of the solid black line. One can see that the tip of the jet quickly reaches its terminal velocity $\simeq v_{\rm d}$, comparable to the bubble velocity $U_{\rm bub}$ (see Fig.~\ref{fig:uterm}) but much smaller than the gas velocity within the jet ($\simeq3U_{\rm bub}$). The increase of the ram pressure with the rise velocity prevents the continuous acceleration of the bubble. Such a velocity corresponds to a Froude number $\simeq1$ and implies that internal gravity waves are efficiently generated by the jet (see \citetalias{Zhang2018} and their fig.~12), providing an important pathway for the uplifted gas to release its energy. In contrast, the gas inside the jet is shielded and moves at a higher velocity. Its motion is similar to a buoyant oscillation.

In our simulations, the reverse jet significantly disturbs the cluster core. It penetrates the cluster center and propagates to a large radius on the opposite side (e.g., $\simeq50\kpc$ in E4L12). We note that, in reality, the interactions of multiple bubbles and their wakes may dramatically change this picture. For example, a symmetric pair of bubbles may lead to a head-on collision of two reverse jets, or the jet may interact with a newly formed bubble in the inner region (particularly relevant to Perseus). Note that the innermost bubbles may expand supersonically while the reverse jets have subsonic velocities. Therefore, such interactions may show a very different physical picture compared to those presented in the last panels in Fig.~\ref{fig:part_a4vb_2}. In addition, 3D instabilities may affect the morphology of the jet, which is not captured in our simulations. In spite of that, our results imply that a large amount of bubble energy is transferred to the uplifted gas (e.g., $\sim50$ per cent of the bubble-released energy goes into the kinetic energy in the bubble wake in E4L12) and further spread into the ICM through turbulence, internal gravity waves, etc. The reverse jet plays an important role in this process and may modulate the central supermassive black hole activity by perturbing the gas core. In the meanwhile, new vortices are formed behind the bubble to replace the detached ones. They usually have a smaller vertical size governed by the buoyancy length scale $\ell_{\parallel}$ (see fig.~4 in \citetalias{Zhang2018}).

\subsection{Effect of Froude number} \label{sec:result:fr}

The Froude number $\Fr\ (\equiv U_{\rm bub}/L_{\rm bub}N_{\rm BV})$ is an essential parameter characterizing the interaction between the bubbles and the ICM. It is important to understand how our picture depends on it. \citetalias{Zhang2018} showed that buoyant, flattened bubbles ($\varepsilon_{\rm bub}\gtrsim4$) tend to have $\Fr\sim1$ (see their fig.~11). For this reason, we will only consider $\Fr$ numbers around this value. To change the bubble's Froude number in our simulations, we drive bubbles that have the same shape and size but different constant rise velocities (i.e., $U_0=75,\ 150$, and $300\kms$). In particular, the bubble with $U_0=150\kms$ is expected to behave similarly to the buoyant case in the run E4L12 (see Fig.~\ref{fig:uterm}). Its corresponding Froude number is $\Fr\simeq1.3$ assuming $N_{\rm BV}=10\,{\rm rad\,Gyr^{-1}}$.

\begin{figure*}
\centering
\includegraphics[width=0.9\linewidth]{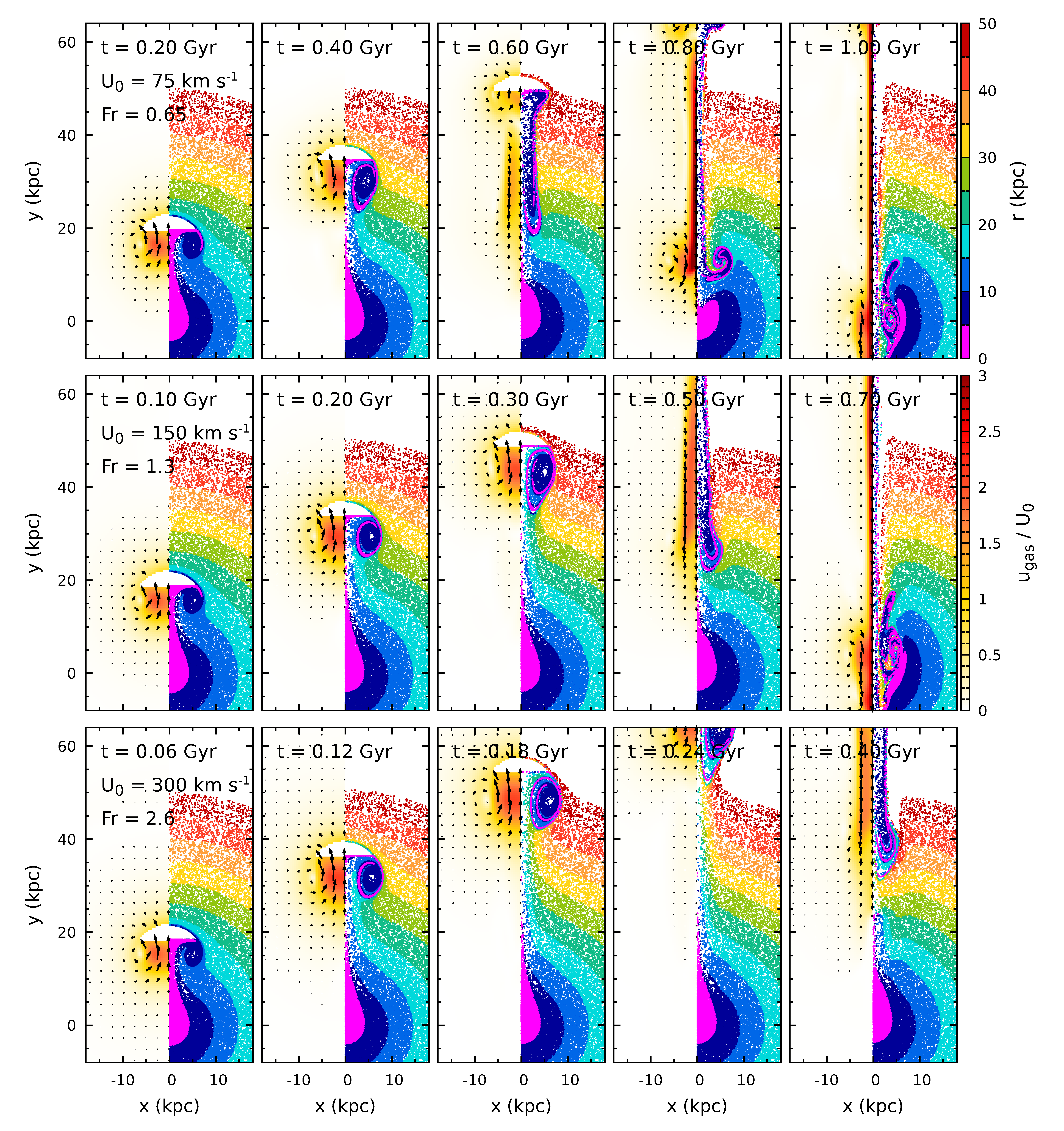}
\caption{Evolution of the gas uplift by bubbles with different Froude numbers (but the same shape and size). The bubbles move with constant velocities: $U_{0}=75,\ 150$, and $300\kms$ in the top to bottom panels (runs E4L12U75/150/300). The Froude number differs by a factor of $2$ between each two adjacent rows. In these simulations, the distribution of the bubble-driven gas velocity shows weak dependence on Fr when scaled by $U_0$, i.e., $u_{\rm gas}/U_0\simeq2$ in the eddies and $\simeq3$ in the reverse jets. The two-phase evolution (i.e., uplift and detachment) takes place in all three cases. The eddy detachment, however, occurs at smaller radii when Fr is smaller (see Section~\ref{sec:result:fr}). }
\label{fig:part_a4fr}
\end{figure*}

Fig.~\ref{fig:part_a4fr} compares simulations with different Froude numbers. In general, their evolution is similar, as discussed in Section~\ref{sec:result:pic}. The two-stage process takes place in all three cases. The normalized gas velocity distribution ($u_{\rm gas}/U_{0}$)
is approximately $\simeq2$ in the bubble eddies and $\simeq3$ in the reverse jets and shows weak dependence on Fr. The moment when the eddies start to detach from their bubble is sensitive to Fr. The reason for this is as follows. The buoyancy length $\ell_{\parallel}\ (=\Fr L_{\rm bub})$ of the system characterizes the length scale at which gravity regulates the growth of eddies. The vertical size of the bubble's primary eddies can be written as $\kappa L_{\rm bub}$. When $\Fr\gg1$, the scaling parameter ranges $\kappa\simeq1-2$, depending only mildly on the bubble's Reynolds number \citep[see, e.g.,][]{Fornberg1988,Lee2000}. When $\ell_{\parallel}<\kappa L_{\rm bub}$ (i.e., $\Fr<\kappa$), gravity dominates and the eddies stretch and detach from the bubble rapidly. Otherwise, the effect of the gravitational force is mild and the eddies maintain their morphology for a longer time. Such a trend is clearly seen in Fig.~\ref{fig:part_a4fr}. It is worth noting that, to show the effect of Fr, we should always compare our snapshots from different simulations at the same time scaled by the eddy turnover time $L_{\rm bub}/U_{\rm bub}$. Then, the bubbles would be located at approximately the same radius. In Fig.~\ref{fig:part_a4fr}, one can see that, at fixed $L_{\rm bub}$, the eddy detachment occurs earlier (i.e., at a smaller radius) the smaller is the bubble's Froude number.

\subsection{Effect of bubble shape and size} \label{sec:result:shape}

\begin{figure}
\centering
\includegraphics[width=0.9\linewidth]{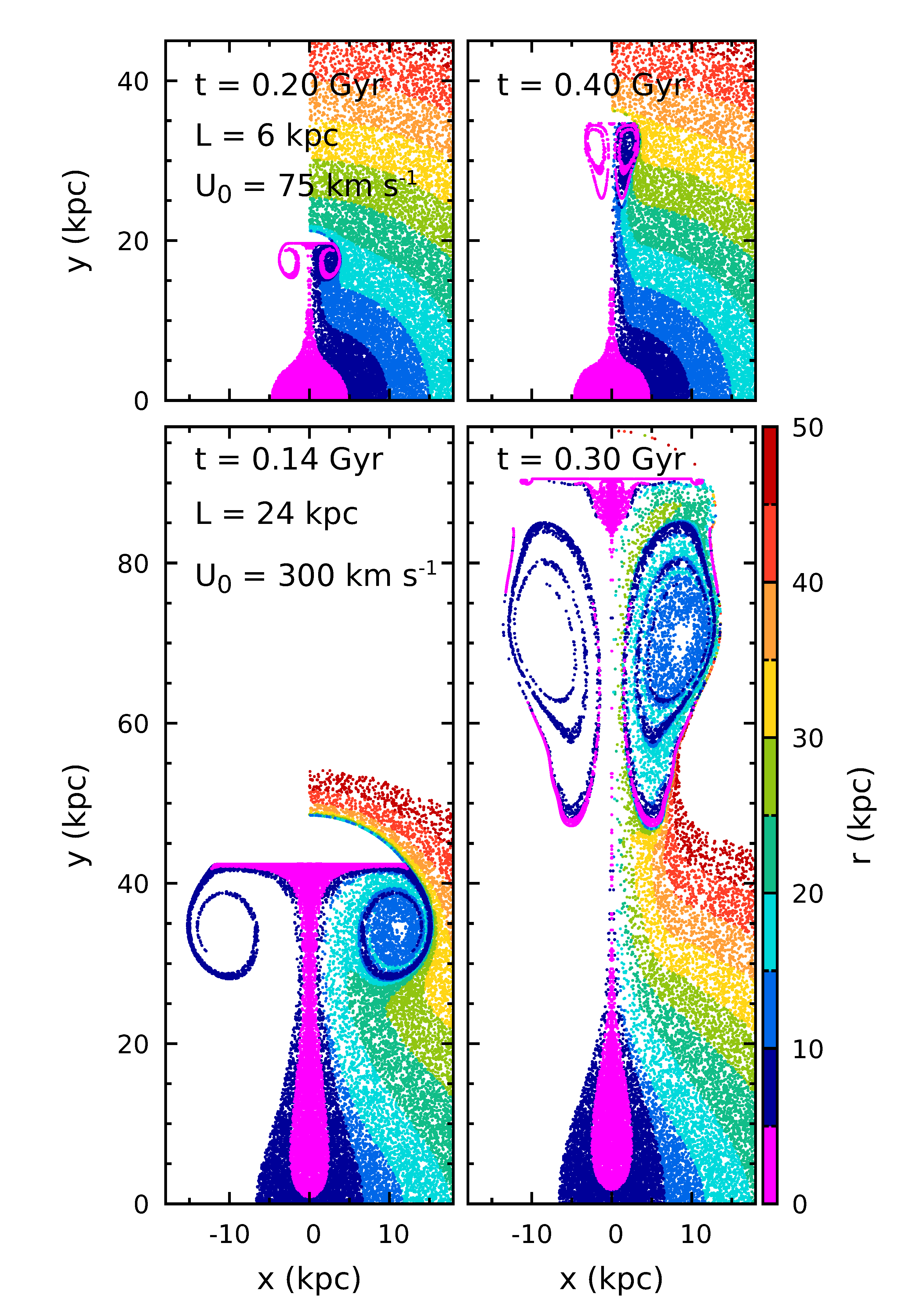}
\caption{A comparison of the simulations E4L6U75 and E4L24U300, where the bubbles have different sizes $L_{\rm bub}=6$ (top) and $24\kpc$ (bottom) but the same shape and Froude number as in the run E4L12U150 (see the middle panels in Fig.~\ref{fig:part_a4fr}). When scaled by $L_{\rm bub}$, these simulations show similar results in terms of the eddy morphology and evolution. More gas is uplifted to a larger cluster radius when $L_{\rm bub}$ is larger (see Section~\ref{sec:result:shape}). }
\label{fig:part_a4l}
\end{figure}

Fig.~\ref{fig:part_a4l} compares our simulations that feature different bubble sizes but the same Froude number as in E4L12U150 (see the middle panels of Fig.~\ref{fig:part_a4fr}). When scaled by $L_{\rm bub}$, they show similar results in terms of the eddy morphology and evolution. This is not surprising because the bubble size in all our simulations is always smaller than the atmosphere's pressure scale height $H_{\rm p}$, which is the essential length scale in our problem. More gas is uplifted to a larger radius when the bubble is larger. We stress here that, in reality, the X-ray bubbles continuously expand while rising in galaxy clusters, with the bubble pressure maintaining equilibrium with the ambient ICM. Our rigid-bubble model cannot capture such a process, but this will only affect the results significantly when $L_{\rm bub}>H_{\rm p}$. Then, the bubble's inflation velocity might become comparable to, or even larger than, the rise velocity.  This, however, is not the case for the Perseus cluster core -- a prototypical example of bubbles in the ICM -- that has $H_{\rm p}\gtrsim50\kpc$ and the outer bubble size $\lesssim20\kpc$.

\begin{figure}
\centering
\includegraphics[width=0.9\linewidth]{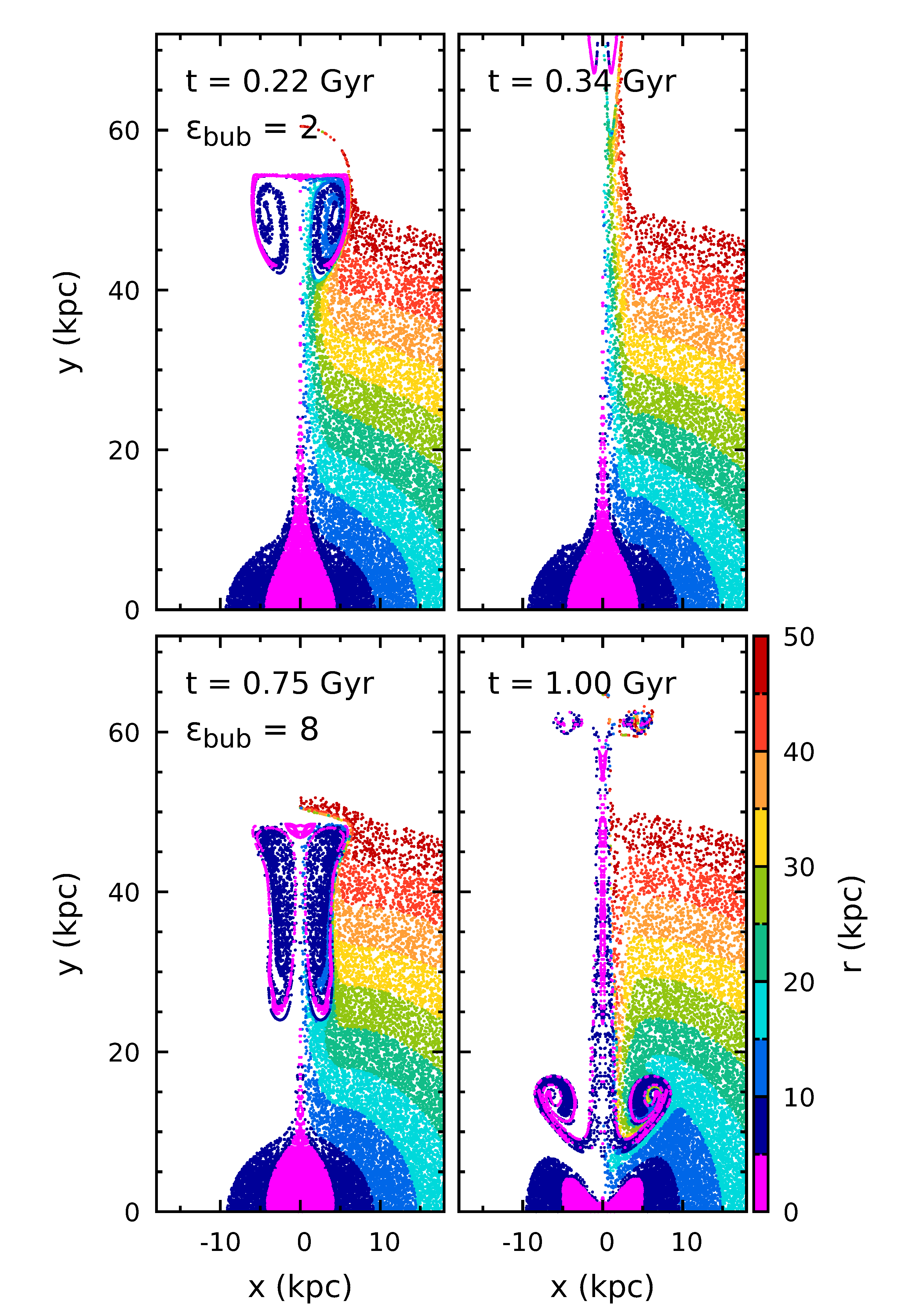}
\caption{Similar to Fig.~\ref{fig:part_a4l} but for bubbles with the same size and different shapes (i.e., aspect ratios $\varepsilon_{\rm bub}$). They rise buoyantly in the simulations E2L12 (top) and E8L12 (bottom). The flatter bubble rises more slowly and, therefore, has a smaller Froude number ($\simeq0.5$). This figure shows a similar result as does Fig.~\ref{fig:part_a4fr}, implying that Fr is a fundamental parameter for our problem (see Section~\ref{sec:result:shape}).  }
\label{fig:part_a4s}
\end{figure}

Fig.~\ref{fig:part_a4s} shows a similar comparison between bubbles with the same $L_{\rm bub}$ but different shapes ($\varepsilon_{\rm bub}=2$ and $8$). While both bubbles buoyantly rise in the simulations, \citetalias{Zhang2018} have shown that bubbles that are flatter along their direction of motion have smaller terminal velocities, corresponding to smaller Froude numbers. Comparing them to the $\varepsilon_{\rm bub}=4$ case (see Figs.~\labelcref{fig:part_a4vb_1,fig:part_a4vb_2}), we find a similar trend as in Fig.~\ref{fig:part_a4fr} -- a larger Froude number leads to a later eddy detachment. This shows again that Fr is the essential parameter in our problem. The bubble shape (aspect ratio) does not contribute additional complexity.

\section{Bubble-driven Gas Motions and their Observational Signatures} \label{sec:obs}

Besides direct imaging, kinematics of uplifted gas is an important probe for bubble-mediated AGN feedback. In this section, we explore observable features of gas motions driven by buoyantly rising bubbles, and how they are linked to the bubble properties.

The two most common ways to measure the LOS velocity of the ICM in observations are: (1) using high-resolution X-ray spectroscopy to measure the Doppler line broadening and shift (see Section~\ref{sec:obs:xray}); and (2) through the dynamics of cold gas measured with optical, near-infrared, and sub-mm observations, under the assumption that all gas phases are dynamically coupled (see Section~\ref{sec:obs:halpha}).

\subsection{X-ray-weighted projected velocity} \label{sec:obs:xray}

\begin{figure}
\centering
\includegraphics[width=1.0\linewidth]{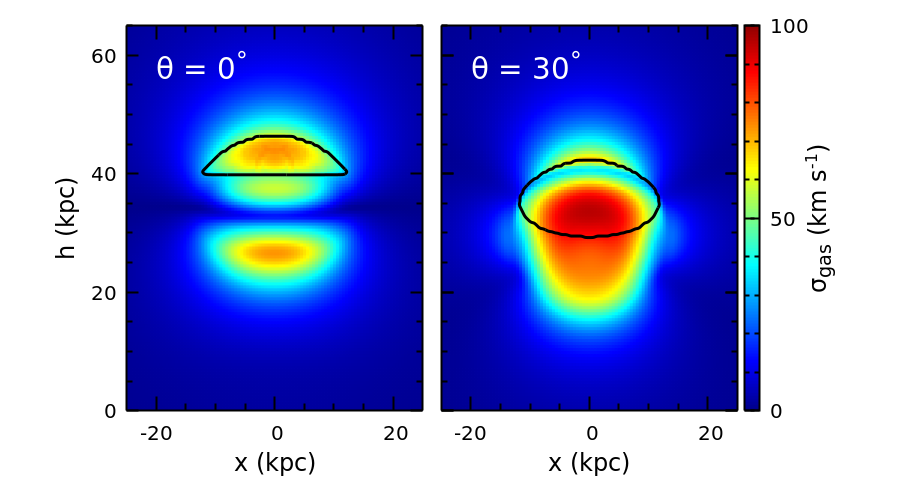}
\caption{Distributions of the X-ray-weighted gas LOS velocity dispersion in the simulation E4L24 at $t=0.3\Gyr$. The left and right panels have the LOS inclination angle $\theta=0$ and $30^{\circ}$ with respect to the inverted $z$-axis, respectively. The black contours indicate the projected outer boundaries of the bubbles. When moving in the plane of the sky (left), the bubble generates a small LOS velocity dispersion ($\simeq3$ times smaller than the bubble-rise velocity) due to the projection effect. When $\theta$ is finite (right), the radial velocity of the uplifted gas contributes dominantly to the observed velocity dispersion (see Section~\ref{sec:obs:xray}). }
\label{fig:usigma}
\end{figure}

Fig.~\ref{fig:usigma} shows the X-ray-weighted LOS velocity dispersion of the gas at $t=0.3\Gyr$ in the simulation E4L24, estimated as
\begin{equation}
  \sigma_{\rm gas}^2 = \int_{\rm LOS}{(u_{\rm gas,l} - \bar{u}_{\rm gas})^2{\epsilon_{\rm gas}\dd\ell}},
\label{eq:usigma}
\end{equation}
where $\bar{u}_{\rm gas}\equiv\int_{\rm LOS}{u_{\rm gas,l}\epsilon_{\rm gas}\dd\ell}$ is the X-ray-weighted LOS mean velocity, and $\epsilon_{\rm gas}\equiv\rho_{\rm gas}^2/\int_{\rm LOS}{\rho_{\rm gas}^2\dd\ell}$ and $u_{\rm gas,l}$ are the normalized X-ray emissivity and gas LOS velocity, respectively. This specific snapshot is selected to match approximately the parameters of the northwestern bubble in Perseus cluster (see Fig.~\ref{fig:perseus}). Its distribution of tracer particles is similar to that in the bottom-left panel in Fig.~\ref{fig:part_a4l}. Due to the axis-symmetry of our 2.5D simulations, we confine the LOS to the $y$--$z$ plane and define the inclination angle $\theta$ as the angle between the LOS and the inverted $z$-axis. Fig.~\ref{fig:usigma} compares the velocity dispersion projected along the LOS, for $\theta=0$ and $30^\circ$. The overlaid black contours show the projected outer boundaries of the rigid bubbles. When the bubble moves in the plane of the sky ($\theta=0$), the velocity dispersion shows two peaks with the maxima $\simeq70\kms$. One peak is at the bubble's projected position, the other is near the rear of the eddies. They show the regions where the uplifted gas has the largest tangential velocity (see Fig.~\ref{fig:part_a4vb_1}). The peak velocity dispersion, however, is $\sim3$ times smaller than the bubble-rise velocity ($\simeq200\kms$), due to the fact that the bubble size ($L_{\rm bub}=24\kpc$) is smaller than the pressure scale height of the cluster core ($\sim50\kpc$; see Fig.~\ref{fig:init_profs}). The projection effect reduces the significance of the bubble-driven velocity dispersion. For the case of non-zero inclination angle ($\theta=30^\circ$), the LOS velocity dispersion is larger with maximum $\simeq100\kms$. It is largely contributed by the bulk motion of the bubble-uplifted gas along the radial direction ($\simeq \sin{\theta}\,U_{\rm bub}$).

\textit{Hitomi} recently measured the gas LOS velocity in the core of the Perseus cluster \citep{Hitomi2018}. It found a mostly uniform velocity dispersion in the cluster core near $\sigma_{\rm obs}\simeq150\kms$ (see their fig.~4). A velocity dispersion excess ($\sigma_{\rm obs}\simeq200\kms$) is detected downstream of the northwestern bubble (which has horseshoe-shaped H$\alpha$ filaments associated with it). Our simulations show that this is likely induced by the wake of the bubble. To interpret the observations, we need to consider two sources contributing to the measured velocity dispersion excess, viz., (1) the ``bulk'' motion component $\sigma_{\rm bulk}$, largely driven by a bubble and/or gas sloshing in cool-core clusters, and (2) $\sigma_{\rm turb}$ due to well-developed turbulence on small scales. If we simply assume that the two components are independent, we have
\begin{equation}
  \sigma_{\rm obs}^2 = \sigma_{\rm bulk}^2 + \sigma_{\rm turb}^2.
\label{eq:obs_sigma}
\end{equation}
In our model, it is reasonable to assume $\sigma_{\rm bulk}\simeq\sigma_{\rm gas}$, fully determined by the simulations, because gas sloshing motions in the Perseus cluster are in general on larger scales than the bubble \citep[see, e.g., ][]{Walker2017}, having only a limited impact on $\sigma_{\rm bulk}$ in the region around the bubble eddies. The second term of Equation~(\ref{eq:obs_sigma}) could be approximately constrained by \textit{Hitomi}'s measurement in the region far from any bubble, $\sigma_{\rm turb}\simeq150\kms$. If that is the case, $\sigma_{\rm gas}\simeq130\kms$ is required to match the excess, which is consistent with our model if $\theta\simeq30-40^{\circ}$ (see Fig.~\ref{fig:usigma}). Note that, $\sigma_{\rm turb}$ and $\sigma_{\rm bulb}$, in fact, cannot be fully independent. One can easily imagine turbulence being stronger close to a bubble. However, given that the bubble size is smaller than the size of the cluster core (e.g., $H_{\rm p}$) in our case, we may only mildly underestimate $\sigma_{\rm turb}$ and thus overestimate $\sigma_{\rm bulb}$ around the bubble region.

\begin{figure*}
\centering
\includegraphics[width=1\linewidth]{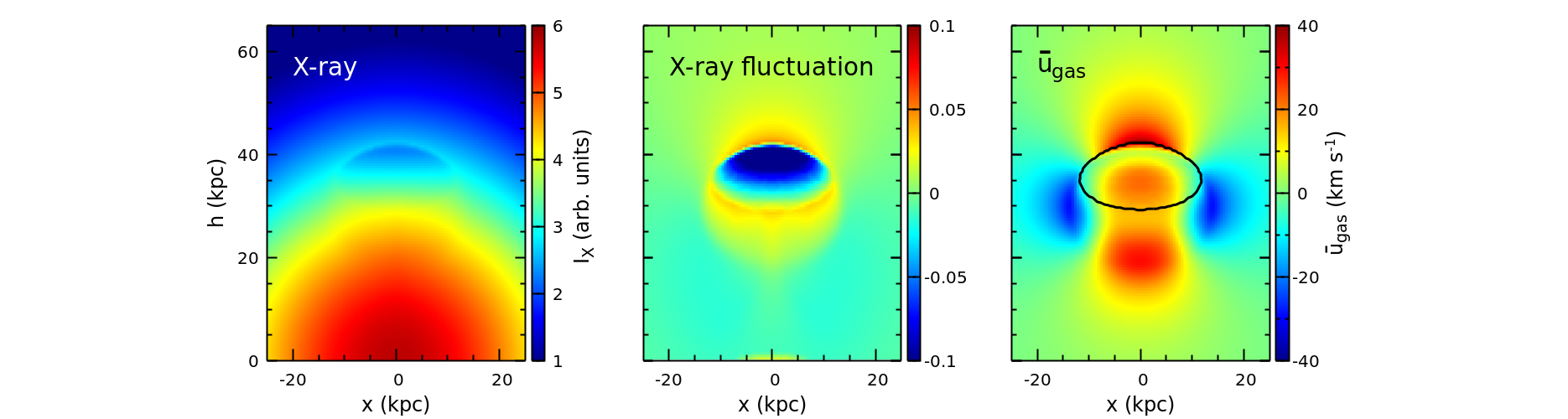}
\caption{Distribution of the X-ray surface brightness (in arbitrary units; left), X-ray fluctuations with respect to the unperturbed atmosphere (middle), and the X-ray-weighted gas LOS mean velocity (right) in the simulation E4L24 with $\theta=30^{\circ}$, corresponding to the right panel in Fig.~\ref{fig:usigma}. The bubble is visually flatter in the X-ray image than its projected outer boundary (black contour), due to its shape being a spherical cap. The LOS mean velocity is positive just downstream of the bubble and negative on the outer sides of the eddies (see Section~\ref{sec:obs:xray}). }
\label{fig:umean}
\end{figure*}

Fig.~\ref{fig:umean} shows the corresponding LOS mean velocity with $\theta=30^{\circ}$ (the right panel), which is positive ($\simeq30\kms$) near the centerline of the bubble and its wake, and negative ($\simeq-30\kms$) on the outer sides of the eddies, generally in line with the \textit{Hitomi} observation as well. The fine velocity structures are expected to be resolved with future high-resolution X-ray observatories (e.g., \textit{Athena}).

There is still one issue that remains -- the projected bubble shape appears overly round when $\theta\simeq30^{\circ}$, compared to the real bubble in the X-ray residual image (see Fig.~\ref{fig:perseus}). Besides the possible overestimation of $\sigma_{\rm bulb}$ discussed above, it could be also partially caused by the bubble shape being a spherical cap. The left panel of Fig.~\ref{fig:umean} shows the distribution of the X-ray surface brightness $I_{\rm X}\propto\int_{\rm LOS}{\rho_{\rm gas}^2\dd\ell}$ of our model with $\theta=30^{\circ}$. The bubble's top half is dimmer than the bottom due to the fact that the plain underside of the bubble is facing towards the observer (see also the middle panel for the X-ray fluctuations). The X-ray observations may also misidentify the bubble boundary due to the complex and non-symmetric gas structures in the inner region of Perseus. It has earlier been shown that the shape of an X-ray cavity is sensitive to the method of extracting a residual image from the total surface brightness map \citep[see, e.g.,][]{Zhuravleva2015}. Yet another possibility is that the velocity excess detected by \textit{Hitomi} is lower in reality (at least with large uncertainties), given that the observations were conducted during the commissioning phase. Future well-calibrated \textit{XRISM} observations will verify the velocity excess in Perseus and observe similar filamentary structures in M87/Virgo \citep{XRISM2020}. We note that the LOS velocities of the H$\alpha$ filaments support the presence of the velocity excess in the bubble downstream (see Section~\ref{sec:obs:halpha}). Finally, it is also plausible that, given the aspect ratio, our simulations underestimate the bubble-rise velocity due to the absence of the bubble-jet interaction and bubble deformation. A flatter bubble may move at $\simeq200\kms$ and would largely ease the discrepancy.

Compared to velocity fields driven by eddies, those associated with the reverse jets are even harder to detect in X-rays, given that the jets are narrow, always subsonic, and not significantly dense (at least in our model without radiative cooling), which do not contribute much to the surface brightness along the LOS. Despite, their detection may provide a unique opportunity to constrain the bubble lifetime and the lower limit of the cluster radius a bubble could reach.

\subsection{LOS velocities of H$\alpha$ filaments} \label{sec:obs:halpha}

Besides X-rays, it is possible to study hot-gas velocity fields using optical H$\alpha$ and sub-mm (e.g., CO) observations of cold gas. The spectral resolution of optical and sub-mm observations is significantly better than that of X-ray telescopes. However, this approach requires an assumption that the two gas phases are efficiently coupled dynamically. Under this assumption, our tracer particles would track the filament motions in the simulations.

\begin{figure*}
\centering
\includegraphics[width=1.0\linewidth]{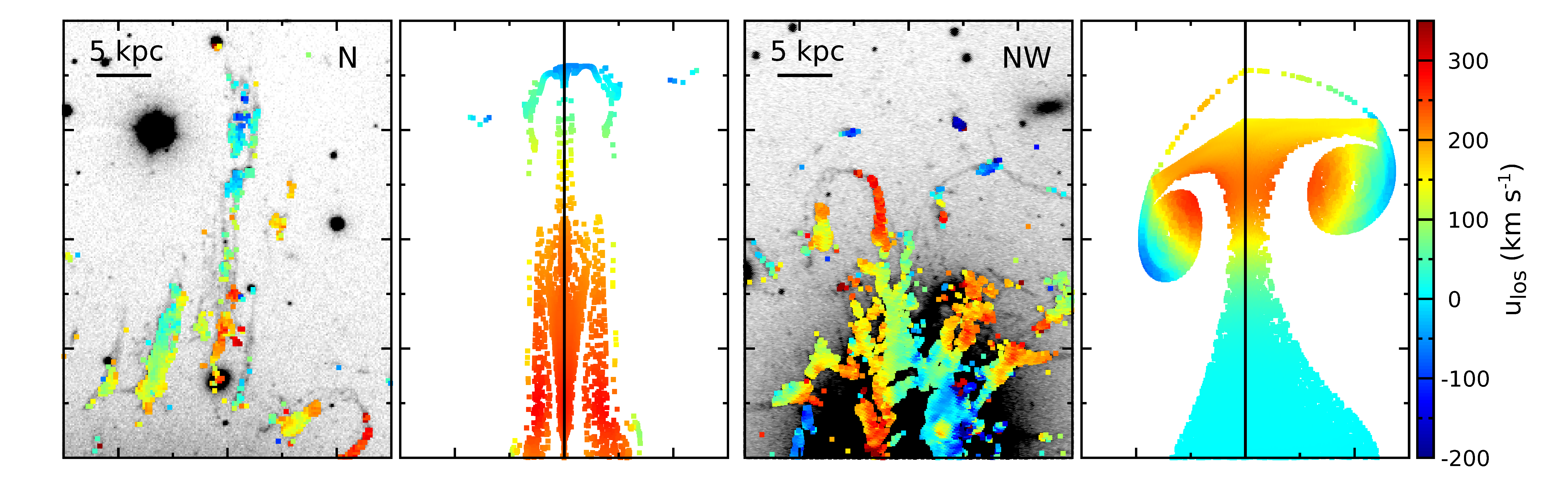}
\vspace{-10pt}
\caption{Model of H$\alpha$ filaments in the Perseus cluster based on the simulation E4L24. For comparison, the real observations are exhibited in the first and third panels, which show the northern (the first panel) and northwestern (the third panel) filaments enclosed by the two white rectangles in Fig.~\ref{fig:perseus}, respectively. They are the two most remarkable ($\sim50\kpc$-long straight and horseshoe-shaped) H$\alpha$ structures in Perseus. The filaments' LOS velocities and surface brightness are shown as the color points \citep{Gendron-Marsolais2018} and the background grey image \citep{Conselice2001}, respectively. The two corresponding numerical models show the projected distributions of the tracer particles along the LOS at $t=0.6$ (the second panel) and $t=0.3\Gyr$ (the fourth panel), respectively. These particles originally reside in the $r<15\kpc$ region at $t=0$. Note that the actual filling factor of the cold filaments is low in the ICM. The left and right halves of these panels show the projected distributions of the particles located in two planar cross-sections, both parallel to the $y$-axis (see more details in Section~\ref{sec:obs:halpha}). The proper LOS directions ($\theta=-30^{\circ}$ and $40^{\circ}$ for the northern and northwestern filaments) are selected to match our models with the observations. This figure illustrates striking dynamical and morphological similarities between our models and observed H$\alpha$-filamentary structures, suggesting that bubble integrity plays an important role in driving the gas velocity field in cluster cores (see Section~\ref{sec:obs:halpha}). }
\label{fig:halpha_comp}
\end{figure*}

Fig.~\ref{fig:halpha_comp} compares our model with the observed H$\alpha$ distributions and their LOS velocities in Perseus enclosed in the white rectangles in Fig.~\ref{fig:perseus}. In the model, we only include particles initially located within $r=15\kpc$, since most of the cold gas is uplifted by the bubbles from the innermost region in our scenario. Given the axis-symmetry of our simulations, the tracer particles have effectively a ring-shaped geometry if viewed in 3D. To model long and narrow structures resembling observed filaments, we assume that the particles are instead confined to a plane, parallel to the $y$-axis. The right halves of the 2nd and 4th panels in Fig.~\ref{fig:halpha_comp} show the case when the filament is in the $x$--$y$ plane ($z=0$ and $x>0$), while in the left halves, the filament is in the plane $x=-z$ ($x<0$), selected to illustrate the projection effect. To match with the observations, we finally plot particles' projected positions on the sky plane by adopting non-zero inclination angles for the LOS (still in the $y$--$z$ plane), viz., $\theta=-30^{\circ}$ and $40^{\circ}$ for the northern and northwestern filaments, respectively. Note that $\theta$ is the only fine-tuned parameter in our model, the rest (e.g., $L_{\rm bub}$, $\varepsilon_{\rm bub}$) are all set based on the observations. Fig.~\ref{fig:halpha_comp} reveals that our models are well matched to the observations in terms of filament morphology, spatial extent, and the LOS velocity distribution. This excellent correspondence suggests that bubble integrity plays an important role in shaping the velocity field of the ICM in cluster cores. This effect is not captured in most of the simulations reported in the literature.

The velocity gradient revealed in the long northern filament (the first panel in Fig.~\ref{fig:halpha_comp}) is well captured in our model. It can be explained by a stretching process taking place when eddies detach from the bubble (see Fig.~\ref{fig:traj}). In the residual X-ray image (Fig.~\ref{fig:perseus}), we tentatively find a bubble candidate to the north of the filament (marked by the arrow at the top), in line with the expectation of our scenario. If confirmed, it will be a detection of a bubble that has maintained its integrity while rising up to $\simeq70\kpc$ away from the cluster center. The bubble has also survived crossing cold fronts formed by a merger-induced gas sloshing \citep[cf.][]{Zuhone2021}. At the same time, our model produces the horseshoe-shaped filaments, which trace the streamlines downstream of the bubble \citep{Fabian2003}. The ``best-fit'' LOS inclination angle $\theta=40^{\circ}$ in our model supports the scenario discussed in Section~\ref{sec:obs:xray} that the northwestern bubble in Perseus moves slightly away from the plane of the sky.

\begin{figure}
\centering
\includegraphics[width=1\linewidth]{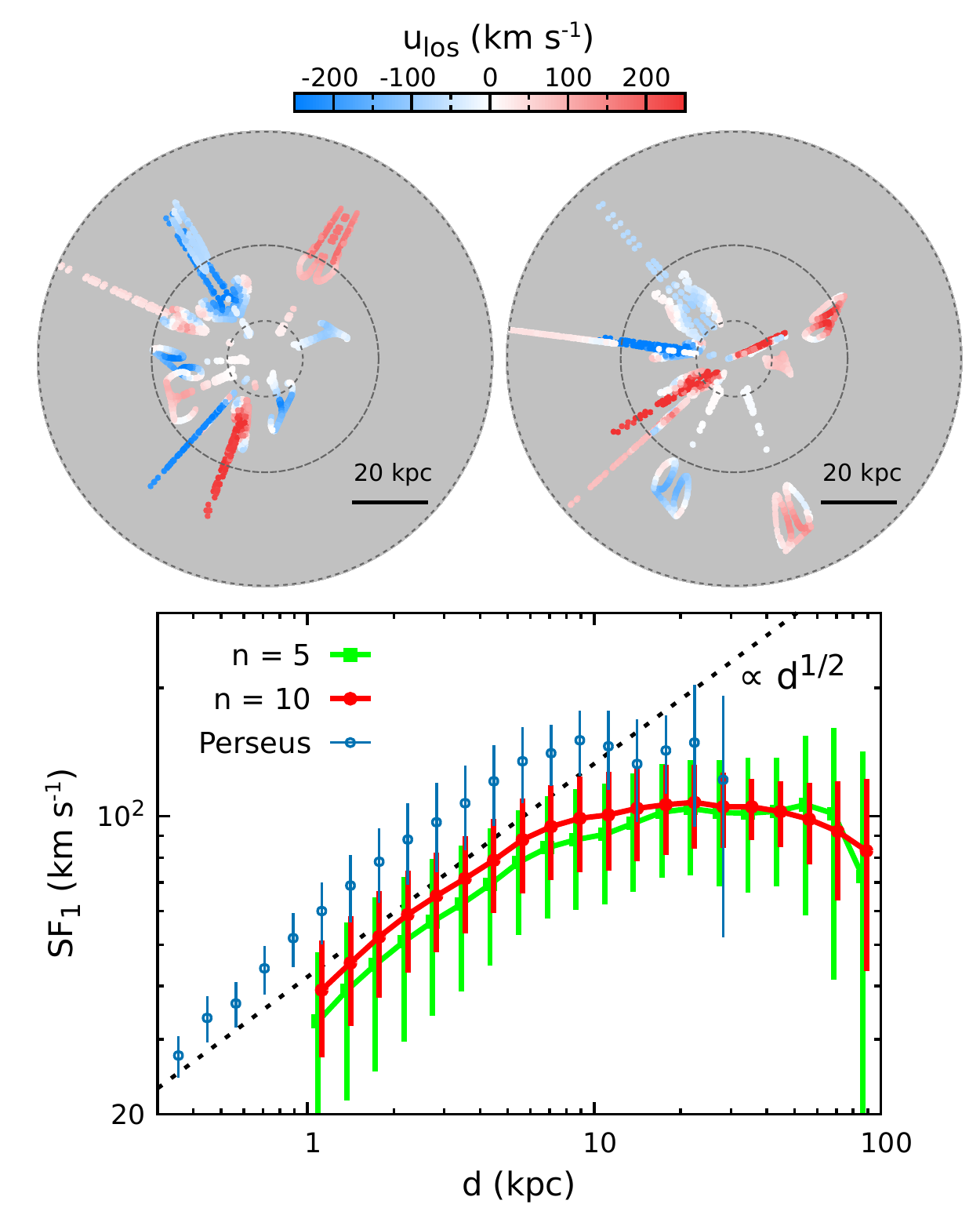}
\caption{\textit{Top panels:} Two examples of the projected distribution of our mock H$\alpha$ filaments, assembled based on the simulation data from the run E4L12S. Three dashed black circles have radii 10, 30, and $60\kpc$ and illustrate the scale of the images. The color encodes the LOS velocity $u_{\rm los}$ of the mock filaments. Each map contains the particles extracted from $n=10$ snapshots that are randomly selected from 13 snapshots spanning from $t=0.2$ to $0.8\Gyr$ (with replacement; see Section~\ref{sec:obs:halpha} for more details on how we generate the mock image). \textit{Bottom panel:} First-order velocity structure function $\rm SF_1$ of our mock filaments. Two curves are produced with $n=10$ (red) and $5$ (green), each averaging over $10^2$ realizations. Error bars indicate the standard deviation. The green curve is slightly shifted leftwards for a clearer view. This figure shows that the shape of our $\rm SF_1$ is close to $d^{1/2}$ (indicated by the dotted black line) on small scales ($d\lesssim10\kpc$) and gradually becomes flat when $d$ is larger. For comparison, the blue points show the observational $\rm SF_1$ of the Perseus cluster (see \citealt{Li2020} for details). To illustrate its random uncertainty, we measure the $\rm SF_1$ in four quadrants separately (see Fig.~\ref{fig:perseus}, where the origin is set at the peak position of the X-rays) and estimate their standard deviation as the error bars. This figure shows that our model agrees well with the observations in both shape and amplitude. This implies an alternative interpretation for the observational data -- the structure function of the H$\alpha$ filaments in cluster cores is dominated by the ``bulk'' velocity patterns driven by intact buoyant bubbles rather than by uniformly distributed turbulence (see Section~\ref{sec:obs:halpha}). }
\label{fig:sf}
\end{figure}

\citet{Li2020} measured the first-order velocity structure function of the LOS velocity of H$\alpha$ filaments in nearby clusters to probe turbulence in the ICM. Here, we carry out a similar exercise for our simulated bubble-driven gas velocity fields. Although small-scale turbulence could not be captured in 2D, our model sheds light on how the ``bulk'' (laminar) motions of the gas within the filaments and large eddies driven by intact bubbles contribute to the velocity structure function, as well as explore the effects of overlapping filaments and their spatial distribution on large scales. For this purpose, we generate a mock distribution of the filaments in 3D. We randomly select $n$ snapshots spanning from $t=0.2$ to $0.8\Gyr$ with replacement in the simulation E4L12S. From each snapshot, we extract one filament constructed from the particles initially located within $r=5\kpc$ (at $t=0$) but within the shell $10\kpc<r<60\kpc$ at the present time~$t$ and assume that the filament resides only in one plane (e.g., the cross section between the ring-shaped particles and the $x-y$ plane). We then set random orientations for these $n$ filaments and assemble them together to form a 2D projected distribution. We assume simply that those filaments do not interfere with each other, though the situation could, of course, be more complicated in reality. The top panels of Fig.~\ref{fig:sf} show two realizations of our mock distributions with $n=10$. The color encodes the filament LOS velocities $u_{\rm los}$. These examples have morphologies that are similar to real observations (see, e.g., Fig.~\ref{fig:perseus}). We then calculate the first-order structure function of the velocity as
\begin{equation}
{\rm SF_1}(d) = \langle|u_{\rm los}({\bf r_1}) - u_{\rm los}({\bf r_2})|\rangle,
\label{eq:sf}
\end{equation}
where $d=|{\bf r_1}-{\bf r_2}|$, and $\langle...\rangle$ denotes the operator of averaging over all velocity pairs.

The bottom panel of Fig.~\ref{fig:sf} shows our structure functions averaged over $100$ realizations. The error bars indicate the standard deviation. The curves with $n=5$ and $10$ are very similar except for the scatter. Our model shows that, in general, the structure function increases with $d$ approximately as $\sim d^{1/2}$ on small scales ($\sim\kpc$) and gradually becomes flat when $d\gtrsim10\kpc$. The transition is smooth and its characteristic scale ($\simeq10\kpc$) is comparable to the bubble size. Qualitatively, one could explain the shape of our $\rm SF_1$ in the following way.
Due to the sparse distribution of the filaments, the structure function is dominated by the velocity pairs across two separate filaments on large scales. If motions of the gas that forms a filament are mostly driven by a bubble, there will be no significant correlation between the filaments. That is exactly the case in our model by design. It explains why the curve tends to be flat in the large-scale regime and why the error bars shrink significantly at $d\gg10\kpc$ when we increase $n$ from $5$ to $10$. In contrast, on small scales, the structure function is largely determined by the velocity gradients in individual filaments.
For the simplest case of a uniformly stretched straight filament, we would have ${\rm SF_1}\propto d$. The structure function's amplitude (but not its shape) depends on the angle between the filament and LOS as well, causing a scatter in the structure function. Meanwhile, substructures in individual filaments, eddies, and overlapping of multiple filaments all complicate the filament distribution on small scales and make the overall slope of the structure function shallower than unity. We can clearly see such complexities in the top panels of Fig.~\ref{fig:sf}. Note that, in our scenario, there is no guarantee that the shape of ${\rm SF_1}$ should be a power law. We have also estimated the structure function for the simulation E4L24 (different bubble size and particle sampling strategy) and found a similar result as in Fig.~\ref{fig:sf}.

The structure function in Fig.~\ref{fig:sf} is surprisingly consistent with the observations by \citet{Li2020} in both amplitude and shape, especially given that no parameter is fine-tuned in our model. This result reveals an alternative interpretation for these observations, namely that, the structure function is dominated by the characteristic velocity pattern and spatial structure of the gas driven by intact bubbles in the cluster core rather than by well-developed turbulence, as proposed in \citet[][see also \citealt{Wang2021,Mohapatra2022}]{Li2020}. We also note that, even though there is no 3D turbulence in our simulations, the gas velocity in the bubble wake (e.g., eddies, reverse jet) is $2$ to $3$ times higher than the bubble-rise velocity (see Fig.~\ref{fig:part_a4fr}), much stronger than the turbulent motions ($\simeq150\kms$) measured with \textit{Hitomi} in Perseus \citep{Hitomi2018}. The filament velocities should therefore always be dominated by the bubble-driven ``bulk'' motions rather than by turbulence. A remarkable consistency between our model and H$\alpha$ observations of the gas velocity distribution in filaments suggests that even though small-scale turbulence is likely developed in reality, it may not affect the gas velocity field strongly, at least not on the scale of the bubble.

\begin{figure}
\centering
\includegraphics[width=1\linewidth]{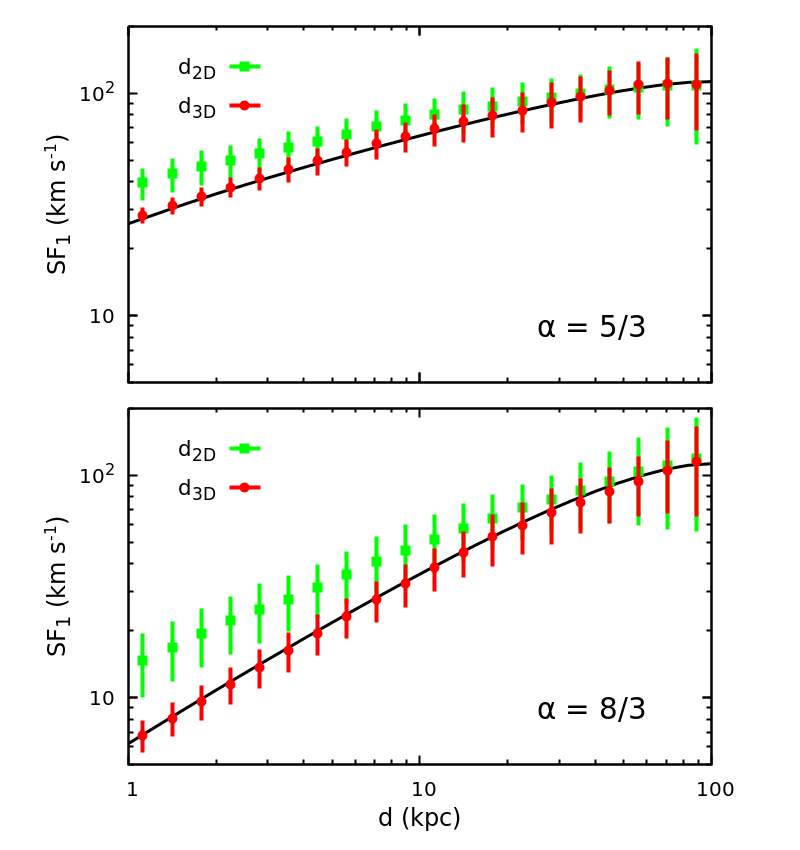}
\caption{Similar to the bottom panel in Fig.~\ref{fig:sf} but the LOS velocity $u_{\rm los}$ of the filaments is replaced by a Gaussian random field in each realization, which is generated with a prescribed underlying energy power spectrum (see Equation~\ref{eq:ps}). In the top panel, the Kolmogorov spectrum $\varpropto k^{-5/3}$ is adopted. In the bottom panel, a steeper spectrum is used to examine if our measurement is sensitive to the spectral index. The green points show the 2D $\rm SF_1$ measured from the projected 2D velocity field (similarly to Fig.~\ref{fig:sf}). The red points show the 3D $\rm SF_1$ using the full spatial information of the filaments. The solid black lines represent the baselines directly estimated from the data cube of the Gaussian random field. This figure shows that, in our setups, the 3D $\rm SF_1$ recovers the input model well; the 2D one, however, underestimates the slope in the inertial range due to projection effects (see Section~\ref{sec:obs:halpha}).}
\label{fig:sf_gauss}
\end{figure}

Since the slope of our structure function is steeper than the prediction for the turbulence in the classical Kolmogorov scenario ($\varpropto d^{1/3}$), it is possible that, on sufficiently small scales, turbulence becomes dominant. It is thus interesting to ask if estimating the velocity structure function for H$\alpha$ filaments is a robust way to detect turbulence properties on those scales. How do projection effects and sparsity of the filaments' distribution affect the measurements? We shed some light on these questions by performing a simple experiment based on our simulations. We generate Gaussian random velocity fields with an underlying energy spectrum
\begin{equation}
E(k) = C_0(k/k_{\rm c})^{-\alpha}e^{-(k_{\rm c}/k)^2},
\label{eq:ps}
\end{equation}
where $C_0$ is a normalization constant setting the standard deviation of the velocity field as $100\kms$, $\alpha$ is the spectral index, $k_{\rm c} = 1/300\kpc^{-1}$ is the cutoff wavenumber fixed at a scale larger than our system. Note that, we are only interested in the small scale ($k\gg k_{\rm c}$) in this experiment. In reality, the characteristic $k_{\rm c}$ should be related with the specific driving mechanisms of the turbulence. The shape of small-$k$ spectral tail might be also shallower (see, e.g., \citealt{Hosking2022}).

We repeat our procedure of generating $n=10$ filaments in 3D, projecting them and calculating the structure function (Fig.~\ref{fig:sf}), but we replace the LOS velocity of the filaments with the Gaussian random field.\footnote{The new filament velocity is proximally interpolated from the data cube of the Gaussian random field, whose spatial resolution is $0.5\kpc$.} For comparison, we also calculate the 3D structure function based on the full spatial information of the filaments rather than the projected one. The results are shown in Fig.~\ref{fig:sf_gauss}, where we examine two different energy spectra with $\alpha=5/3$ (Kolmogorov type) and $\alpha=8/3$. The solid black lines show the baseline results directly estimated from the 3D data cube of the velocity field. Our 3D $\rm SF_1$ (the red lines) recovers the input accurately for both spectra. The sparsity of the filaments does not have any effect on the measurement. The 2D $\rm SF_1$, however, obviously biases the curves due to projection effects \citep{Li2020}. It infers a shallower energy spectrum in the inertial range compared to the baseline model. However, careful modeling (e.g., as done in Fig.~\ref{fig:sf_gauss}) might be used to correct this bias. We conclude that measuring structure functions of the embedded cold gas may still be a robust approach to detect turbulence in the ICM, as long as (1) the cold and hot gas phases are tightly coupled together and (2) the gas velocity is dominated by turbulent motion.

\section{Conclusions} \label{sec:conclusions}

In this study, we start with an assumption that buoyant bubbles in cluster cores maintain their shapes during their rise. We model them as cap-shaped bodies and show the important role of such ``rigid'' bubbles and their long-term interactions with the ambient ICM in shaping the gas velocity distribution and forming thin gaseous structures (e.g., H$\alpha$ filaments) in cluster cores.

In our rigid-bubble simulations, in the wakes of buoyantly rising bubbles, eddies are formed with a size comparable to the bubble. The morphology of their streamlines is similar to the horseshoe-shaped H$\alpha$ filaments observed in the Perseus cluster, supporting the scenario proposed by \citet{Fabian2003}. We find that most of the gas mass uplifted by bubbles is through “eddy transport” rather than Darwin drift, i.e., the gas is trapped inside the downstream eddies and moves together with the rising bubble. The evolution of the eddies is significantly affected by the gravitational field during the rise, characterized by the bubble's Froude number $\Fr$. The eddies are gradually elongated along the bubble's direction of movement and eventually detach from the bubble. In this process, the stretched eddies quickly evolve into a high-speed reverse jet-like stream propagating towards (and possibly even shooting through) the cluster center in our model. The typical gas velocity in the bubble wake is higher than the bubble-rise velocity, e.g., by a factor of $\simeq2$ in the eddies and $\simeq3$ in the jet, showing only weak dependence on Fr (see Fig.~\ref{fig:part_a4fr}). The jet structure provides a natural explanation for the $\sim10^2\kpc$ long cold gas filaments observed in nearby clusters (e.g., Perseus; see Fig.~\ref{fig:perseus}).

We make a detailed comparison of our model with the observations of the Perseus cluster using both the X-ray-weighted gas velocity dispersion \citep{Hitomi2018} and LOS velocities of H$\alpha$ filaments \citep{Gendron-Marsolais2018}. Using our simulations, we predict the terminal velocity of the northwestern bubble in Perseus ($\simeq200\kms$; see also \citetalias{Zhang2018}) and find that an inclination angle $\theta\simeq30^\circ$ of the LOS is required to explain the velocity dispersion excess ($\simeq200\kms$) observed downstream of the bubble by \textit{Hitomi} (see Fig.~\ref{fig:usigma}). If the bubble moved in the plane of the sky, the observed peak velocity dispersion would be $\sim3$ times smaller than the bubble-rise velocity, due to the fact that the bubble size is smaller than the scale of the cluster’s gas core.

Under the assumption that the cold and hot gas phases efficiently couple dynamically, our simulations illustrate the formation of both the long straight H$\alpha$ filament and the horseshoe-shaped filament in Perseus. The model matches well the observed filament morphology, spatial extent, and the LOS velocity distribution (see Fig.~\ref{fig:halpha_comp}). To complement this morphological study, we assemble mock distributions of the filaments in 3D to estimate their projected LOS velocity structure function as done in \citet{Li2020}. Without fine-tuning of any parameters, our model reproduces both the observed structure function's amplitude and shape (see Fig.~\ref{fig:sf}), revealing a simple interpretation for the observational measurements -- the structure function of the filaments is dominated by velocity gradients of inward and outward laminar gas flows and large eddies formed behind the bubbles. The uniform, small-scale turbulence, if present, cannot be probed through the structure function of H$\alpha$ filaments because the ``bulk'' velocity pattern is dominant. Overlapping of  filaments on small scales and their sparse distribution on large scales also affect the shape of the structure function at small and large separations, respectively.

Finally, we acknowledge the major simplification made in this study. Given the 2.5D rigid-bubble model adopted by us, (1) bubble deformation during the rise, (2) small-scale 3D turbulence, and (3) rapid inflation of the bubbles in the early phase of the AGN feedback (i.e., bubble-jet interaction) are not captured. Gas cooling is also neglected along with ambient gas flows driven by sloshing. Determining how these processes affect our results quantitatively needs a more detailed investigation in the future. Nevertheless, our model appears to capture several key features of the gas velocity distribution driven by intact rising bubbles in the cluster cores and may help interpret current and future observations (from, e.g., \textit{XRISM}, \textit{Athena}).

\section*{Acknowledgments}

Support for this work was provided by the National Aeronautics and Space Administration through Chandra Award Number TM1-22008X issued by the Chandra X-ray Center, which is operated by the Smithsonian Astrophysical Observatory for and on behalf of the National Aeronautics Space Administration under contract NAS8-03060. Part of the simulations presented in this paper were carried out using the Midway computing cluster provided by the University of Chicago Research Computing Center. IZ is partially supported by a Clare Boothe Luce Professorship from the Henry Luce Foundation. The work of AAS was supported in part by UK EPSRC grant EP/R034737/1. WF acknowledges support from the Smithsonian Institution, the Chandra High Resolution Camera Project through NASA contract NAS8-03060, and NASA Grants 80NSSC19K0116, GO1-22132X, and GO9-20109X.

\section*{Data Availability}

The data underlying this article will be shared on reasonable request to the corresponding author.

\begin{figure*}
\centering
\includegraphics[width=0.9\linewidth]{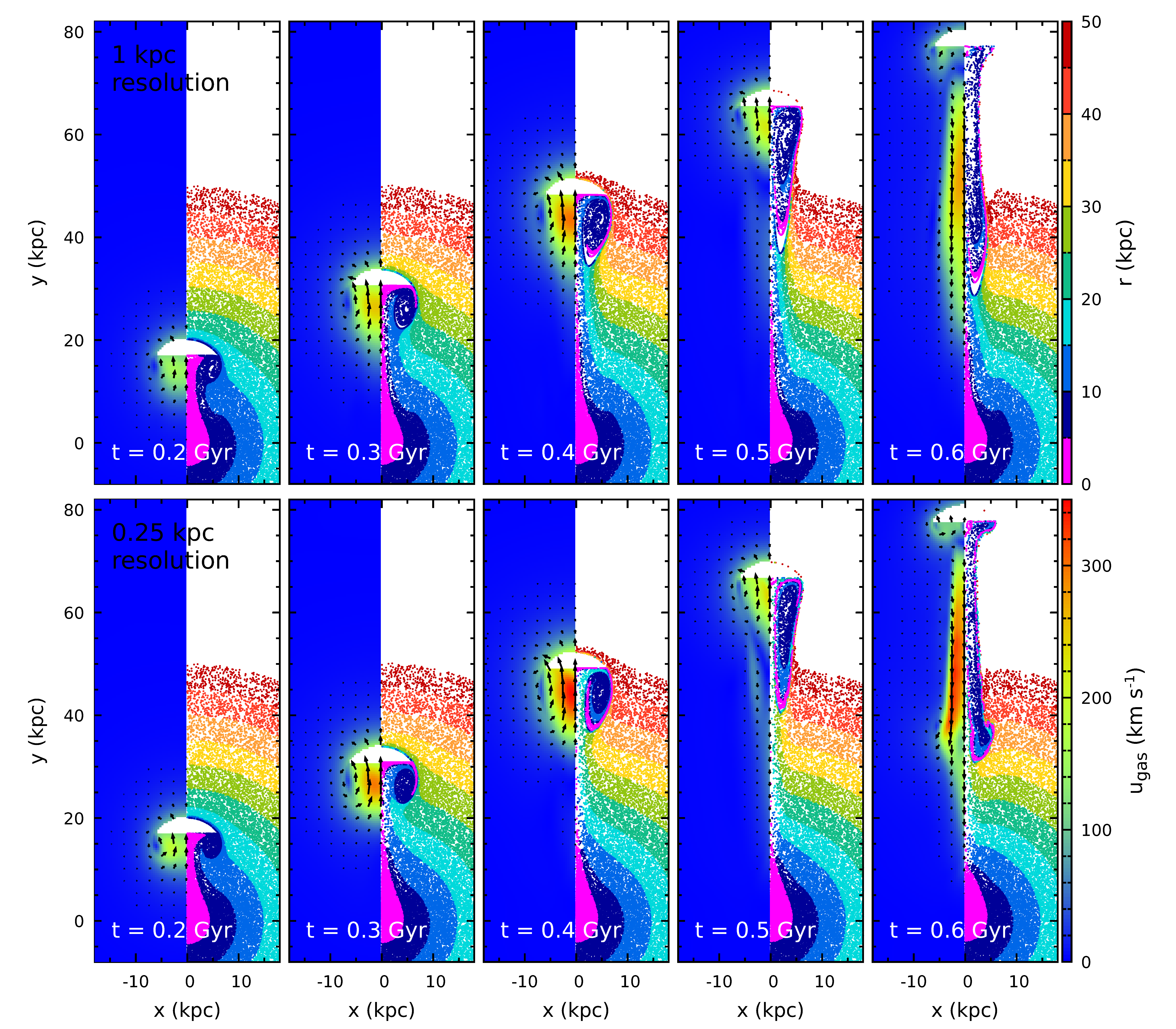}
\caption{A comparison of simulations with different spatial resolutions: $1\kpc$ (top panels) and $0.25\kpc$ (bottom panels). Except for the resolution, they share the same parameters as the run E4L12 (see Fig.~\ref{fig:part_a4vb_1}). The left and right halves of the panels show distributions of the gas velocity and tracer particles, respectively. This figure shows that the resolution does not affect the evolution of the uplifted gas and the global velocity pattern. The resolution mildly affects the velocity amplitude within the eddies and reverse jets (see Appendix~\ref{sec:appendix}). }
\label{fig:part_a4res}
\end{figure*}

\appendix

\section{Resolution test} \label{sec:appendix}

To test how our simulation results depend on the effective spatial resolution, we rerun the simulation E4L12 with two different resolutions, $1$ and $0.25\kpc$, i.e., lower and higher than the default value by a factor of 2. The comparisons of gas velocities and distributions of tracer particles are shown in Fig.~\ref{fig:part_a4res} (see also Fig.~\ref{fig:part_a4vb_1}). The runs with different resolutions show similar results, including the evolution of bubble velocity pattern, formation of eddies, the moment of the eddy detachment, and the formation of a reverse jet. It demonstrates that our findings are not affected by the resolution. Despite, there are minor differences between the low and high-resolution runs. The high-resolution results show stronger gas velocities in the eddies and reverse jet by $\sim10$ per cent and also more prominent instabilities formed near the tip of the reverse jet. These differences are in line with the expectations -- simulations with higher resolution capture more instabilities and turbulent gas structures. These, however, do not affect the velocity structure function modeled in Section~\ref{sec:obs:halpha}.

\bsp	% typesetting comment
\label{lastpage}
\end{document}